\documentclass{aa} 

\usepackage{graphicx} 
\usepackage{amsmath}
\usepackage{mathtools}
\usepackage{amsfonts} 
\usepackage{xcolor}
\usepackage{txfonts}
\usepackage[normalem]{ulem}
\usepackage{lipsum}
\usepackage{subcaption} 
\usepackage{natbib}
\usepackage{hyperref}
\usepackage{float}
\hypersetup{
    colorlinks=true,
    citecolor=blue
}

\newcommand{\sv}{\mathbf{s}}
\newcommand{\kv}{\mathbf{k}}
\newcommand{\xv}{\mathbf{x}}
\newcommand{\rv}{\mathbf{r}}
\newcommand{\nv}{\hat{\mathbf{n}}}

\renewcommand{\vec}{\mathbf}

\newcommand{\bdm}{\begin{displaymath}}
\newcommand{\edm}{\end{displaymath}}

\newcommand{\qv}{\vec{q}}

\newcommand{\zv}{\vec{z}}

\newcommand{\mycomment}[1]{}

\usepackage{lscape}             
\usepackage{placeins}           
                                
\begin{document}

   \title{First full-shape joint analysis of the two- and three-point 
correlation functions on real data: $\Lambda$CDM cosmological constraints from BOSS DR12}

   \titlerunning{Full-shape 2PCF and 3PCF cosmological analysis of BOSS DR12 }


\author{
M. Guidi\thanks{\email{massimo.guidi6@unibo.it}}\inst{1, 2}
\and M. Moresco\inst{1, 2}
\and A. Veropalumbo\inst{3, 4, 5}
\and L. Cavazzini\inst{1}
\and A. Farina\inst{3, 4}
\and A. Labate\inst{1, 2}
}

\institute{
Dipartimento di Fisica e Astronomia, Universit\`a di Bologna,
Via Gobetti 93/2, 40129 Bologna, Italy
\and
INAF–Osservatorio di Astrofisica e Scienza dello Spazio di Bologna, Via Piero Gobetti 93/3, I-40129 Bologna, Italy
\and
INAF-Osservatorio Astronomico di Brera, Via Brera 28, I-20122 Milano, Italy
\and
INFN-Sezione di Genova,
Via Dodecaneso 33, 16146 Genova, Italy
\and
Dipartimento di Fisica, Universit\`a di Genova,
Via Dodecaneso 33, 16146 Genova, Italy   
}

 \abstract{
The three-point correlation function (3PCF) encodes cosmological information 
beyond the two-point correlation function (2PCF), yet a full-shape joint 
analysis in redshift space using real data has so far been lacking. 
In this work, we present the first full-shape cosmological constraints 
from a joint analysis of 2PCF and 3PCF in redshift space, 
using BOSS DR12 data, extending to real data,
including Alcock--Paczy\'nski and redsfhit space distortions, the full-shape 
configuration-space framework validated in real space for the first time by 
\citet{EuclidGuidiEtAl2025}. 
We model both statistics adopting the velocity difference generating 
function (VDG) framework, incorporating non-perturbative Fingers-of-God 
damping, a complete Eulerian galaxy bias expansion, and infrared 
resummation. 
Fast and accurate theoretical predictions are obtained using dedicated 
emulators, which enable a full-shape likelihood analysis of the 3PCF 
and its combination with the 2PCF, varying the cosmological parameters $10^9 A_s, \omega_{\rm cdm}$ and $h$, while the baryon fraction density $\omega_{b}$ is fixed to its fiducial value. The covariance matrix is estimated 
from 2048 MultiDark-Patchy mocks, and an optimal data-vector compression 
ensures a stable covariance inversion. The perturbative model is validated 
against goodness-of-fit tests and posterior stability diagnostics across 
different scales, and provides a good description of the joint data 
vector down to $r_{\rm min}^{\rm 3PCF} \sim 60\,h^{-1}{\rm Mpc}$. 
We find that the joint 2PCF+3PCF analysis yields significant improvements over the 
2PCF-only baseline, with gains of approximately 29\%, 10\%, and 24\% 
on $\sigma(h)$, $\sigma(\omega_{\rm cdm})$, and $\sigma(A_s)$, 
respectively. The improvements found in this analysis mainly arise from the BAO cosmological information 
encoded in the 3PCF triangle configurations, which brings additional information with respect to  the 2PCF alone.
}

   \maketitle
\nolinenumbers

\section{Introduction}
\label{sec:introduction}
Galaxy clustering has emerged as one of the most powerful probes of the
large-scale structure (LSS) of the Universe, offering a direct window onto
the statistical properties of the matter and galaxy density field across a wide range of
scales and cosmic epochs. Over the past two decades, spectroscopic surveys
such as the Baryon Oscillation Spectroscopic Survey
\citep[BOSS;][]{DawsonEtal2013, AlamEtal2017}, the Dark Energy
Spectroscopic Instrument \citep[DESI;][]{AghamousaEtal2016} and \textit{Euclid} \citep{MellierEtal2024} have mapped and are still mapping the
three-dimensional distribution of millions of galaxies, enabling precise
measurements of the two-point correlation function (2PCF) and its
Fourier-space counterpart, the power spectrum. These measurements have
been exploited to use baryon acoustic oscillations (BAOs) as a standard ruler
to reconstruct the background expansion history
\citep{SeoEisenstein2003, EisensteinEtal2005B, BeutlerEtal2014,
RossEtal2007, AdameEtal2024DESI_FS}. They have also been used to probe the growth of cosmic
structures through redshift-space distortions
\citep[RSD;][]{PeacockEtal2001, GuzzoEtal2008, BeutlerEtal2017B, SanchezEtal2017b}, as well as through full-shape analyses
of their redshift-space multipoles
\citep{SanchezEtal2013, DAmicoEtal2020, IvanovSimonovicZaldarriaga2020,
PhilcoxIvanov2022}.

While two-point statistics are sufficient to characterise a Gaussian random
field, the observed galaxy density field deviates significantly from
Gaussianity. This non-Gaussian character arises from multiple sources: the
nonlinear evolution of density perturbations
\citep{Fry1984, BernardeauEtal2002}, the non-linear and non-local relation
between galaxies and the underlying matter field
\citep[galaxy bias;][]{FryGaztanaga1993, Fry1994B,
ChanScoccimarroSheth2012, BaldaufEtal2012,
EggemeierScoccimarroSmith2019}, and redsfhit space distortions (RSD)
\citep{HivonEtal1995, ScoccimarroCouchmanFrieman1999}. Additionally,
primordial non-Gaussianities, potentially generated during inflation, can
leave detectable imprints on the large-scale galaxy distribution
\citep{VerdeEtal2000, Scoccimarro2000A,
ScoccimarroSefusattiZaldarriaga2004}. All these effects transfer
cosmologically relevant information from the two-point to higher-order
statistics, in particular to the three-point correlation function (3PCF) in
configuration space and the bispectrum in Fourier space. The joint analysis of two- and three-point statistics has been shown to 
break parameter degeneracies and significantly tighten cosmological 
constraints \citep[see e.g.][]{SefusattiKomatsu2007, MorescoEtAl2014,  MorescoEtal2017,
DAmicoEtal2020, PhilcoxIvanov2022, DAmicoEtal2022A, CabassEtal2022, 
VeropalumboEtal2021, KuruvillaPorciani2020, MoradinezhadDizgahEtal2021, FarinaEtal2024a,
PugnoEtAl2024a, EuclidGuidiEtAl2025}.

The majority of full-shape analyses exploiting three-point statistics have so
far focused on the bispectrum in Fourier space
\citep{GilMarinEtal2017, DAmicoEtal2020, DAmicoEtal2022A, CabassEtal2022,
PhilcoxIvanov2022, AdameEtal2024DESI_FS, DESINovellMasotEtAl2025, DESINovellMasotEtAl2026, ForeroSanchezEtAl2026}. The preference for Fourier space
is driven by the availability of efficient estimators and well-developed
perturbation theory (PT) frameworks, including Eulerian SPT and Lagrangian PT
\citep{BernardeauEtal2002, Matsubara2008A, Matsubara2008Berr, CarlsonReidWhite2013, WangReidWhite2014}, and the effective field theory of
large-scale structure \citep[EFTofLSS;][]{BaumannEtal2012,
CarrascoHertzbergSenatore2012}. Configuration-space analyses, on the other
hand, offer a natural advantage in handling the complex geometry of
spectroscopic surveys: the Szapudi--Szalay estimator \citep{SzapudiSzalay1997}
automatically accounts for boundary and selection effects, avoiding the need to
convolve theoretical models with survey window functions. The
main obstacles to exploiting the 3PCF have historically been the
$\mathcal{O}(N^3)$ computational cost of direct triplet counting and the
limited availability of accurate redshift-space models. The introduction of a
spherical harmonic decomposition estimator \citep{SlepianEisenstein2015B,
SlepianEisenstein2018} reduced the computational scaling to $\mathcal{O}(N^2)$,
enabling the analysis of catalogues with $N \sim \mathcal{O}(10^7)$ galaxies,
and triggered a rapid development of independent measurement tools
\citep{MarulliVeropalumboMoresco2016, FarinaEtal2024a, EuclidVeropalumbo2026, 2021arXiv210508722P}.
This progress culminated in the first detection of the BAO peak in the galaxy
3PCF BOSS data \citep{SlepianEtal2017}, DESI data \citep{DESIKamalinejad2026} and cluster of galaxies \citep{MorescoEtal2021}, as well as
joint 2PCF+3PCF analyses for the measurement of the linear growth rate of cosmic structures, as shown in 
\cite{VeropalumboEtal2021}.

Advances in 3PCF estimation have driven parallel progress in theoretical
modelling. Perturbative predictions at leading (LO) next-to-leading (NLO) orders for the
real-space 3PCF, incorporating non-local bias and infrared (IR) resummation,
have been validated against $N$-body simulations and synthetic galaxy
catalogues \citep{VeropalumboEtal2022, GuidiEtal2023}. \cite{EuclidGuidiEtAl2025} extended this modelling framework for the first time to a full-shape joint 2PCF+3PCF analysis in real space on
catalogues mimicking the \textit{Euclid} spectroscopic sample, demonstrating that
perturbative models can accurately recover cosmological parameters from
configuration-space data. However, a complete extension to redshift-space data has remained a
critical missing step, primarily due to the absence of accurate and
efficient theoretical predictions for the redshift-space 3PCF suitable
for full-shape likelihood analyses.

In this paper, we present the first full-shape joint analysis of the 2PCF and
3PCF multipoles in redshift space using BOSS DR12 data
\citep{AlamEtal2017}, splitting the sample into two effective redshift bins,
\textit{low-z} ($\bar{z} \simeq 0.38$) and \textit{high-z}
($\bar{z} \simeq 0.61$), obtained via redshift resampling of the original
LOWZ and CMASS galaxy samples.

The theoretical model is built on the velocity-difference
generating function (VDG) framework \citep{EggemeierEtAl2025}, which
provides a non-perturbative treatment of LOS velocity statistics and
their damping effects, combined with a one-loop SPT description of the
density field and a complete galaxy bias expansion including non-local and
higher-derivative terms. The Infrared resummation is implemented following \citet{BlasEtal2016} to
accurately capture the BAO feature in both statistics. Together, these ingredients constitute a complete perturbative model for the configuration-space galaxy clustering signal, directly applicable to
upcoming Stage IV surveys.

This paper is organised as follows. In Sect.~\ref{sec:theory_model} we introduce
the theoretical framework, including the definitions of the 2PCF and 3PCF
multipoles, the VDG model, the galaxy bias expansion, and IR resummation.
Section~\ref{sec:dataset} describes the BOSS DR12 data, the mock catalogues
used for covariance estimation while the estimators and the clustering measurements are given in Appendix \ref{app:estimators}. Section ~\ref{sec:inference} details the likelihood and inference setup.
Finally, our results are presented in Sects. \ref{sec:2pcf_analysis}, \ref{sec:joint_analysis}, where we discuss
the full-shape constraints on cosmological and bias parameters from the 2PCF alone and from the joint 2PCF+3PCF analysis. We draw our conclusions in Sect.~\ref{sec:conclusions}.



\section{Theoretical models}
\label{sec:theory_model}
\subsection{The velocity difference generating function framework}
\label{sec:vdg_framework}
In spectroscopic galaxy surveys, the line-of-sight positions of galaxies 
are inferred from their redshifts, which include a contribution from 
peculiar velocities in addition to the Hubble flow. This induces a 
systematic displacement of galaxies along the line of sight in the 
inferred coordinate frame, the so-called redshift-space distortions 
(RSD), which must be carefully modelled to extract unbiased cosmological 
information from the observed clustering signal. In the distant-observer, 
plane-parallel approximation, the mapping from real-space positions $\xv$ 
to redshift-space positions $\sv$ is written asn as
\begin{equation}
  \sv = \xv + f\,u_{\hat{n}}(\xv)\,\nv\,,
  \label{eq:rsd_mapping}
\end{equation}
where $f \equiv \mathrm{d}\ln D/\mathrm{d}\ln a$ is the logarithmic growth rate, $D$ is the linear growth factor,
$\nv$ denotes the (fixed) LOS, and the LOS velocity field is expressed as
$v_{\hat{n}}(\xv)= f\,\mathcal{H}\,u_{\hat{n}}(\xv)$ with $\mathcal{H}^{-1}$ the comoving Hubble scale. 

The Fourier-space galaxy density contrast in redshift space can be written exactly as \citep{EggemeierEtAl2025}
\begin{equation}
  \delta_s(\kv) = \int_{\xv} e^{i\kv\cdot\xv}\,e^{\lambda u_{\hat{n}}(\xv)}
  \Big[\delta_g(\xv) + f\,\partial_{\hat{n}} u_{\hat{n}}(\xv)\Big]\,,
  \label{eq:delta_s_exact}
\end{equation}
where we introduced the shorthand $\lambda \equiv i f k_{\hat{n}}$, and the term in square brackets reduces to the Kaiser limit at large scales when the exponential is set to unity. Here $\delta_g$ 
denotes the galaxy overdensity field, which encodes the non-linear and non-local 
relation between the observed galaxy distribution and the underlying matter field 
through a galaxy bias expansion; the explicit parametrisation adopted in this work 
is introduced in Sect.~\ref{sec:perturbative_expansion}.

A conventional approach is to expand the exponential in Eq.~\eqref{eq:delta_s_exact} and express $\delta_s$ in terms of redshift-space PT kernels $Z_n$, 
\begin{align}
  \delta_s(\kv) = &  \sum_{n=1}^{\infty}\int_{\qv_1}\cdots\int_{\qv_n}
  (2\pi)^3\delta_\mathrm{D}\!\Big(\kv-\qv_{1\cdots n}\Big)\,
  Z_n(\qv_1,\ldots,\qv_n)\, \notag \\ & \times \delta_{\rm lin}(\qv_1)\cdots\delta_{\rm lin}(\qv_n)\,,
  \label{eq:delta_s_Zn}
\end{align}
which yields the familiar SPT expressions for $P_s$ and $B_s$ when truncating at a fixed order in perturbations.  For a detailed description of PT kernels, see Appendix \ref{app:kernels}. 

In this work, we instead exploit that the mapping in Eq.~\eqref{eq:delta_s_exact} allows an exact reformulation of the
two- and three-point statistics in terms of configuration-space correlators involving exponentials of velocity
differences.  In particular, for the power spectrum one can write \citep{EggemeierEtAl2025}
\begin{equation}
  P_s(\kv) = \int \mathrm{d}^3 r\,e^{i\kv\cdot\rv}\,
  \Big\langle e^{\lambda\,\Delta u_z}\, D(\xv)\,D(\xv')\Big\rangle\,
  \label{eq:Pk_config_cumulant}
\end{equation}
where $\Delta u  \equiv u_{\hat{n}}(\xv)-u_{\hat{n}}(\xv')$, $\rv \equiv \xv-\xv'$ and $D \equiv \delta_g + f\,\partial_{\hat{n}} u_{\hat{n}}$ collects the Kaiser-limit density contribution but evaluated using fully
non-linear fields

Applying a cumulant expansion to the correlator in Eq.~\eqref{eq:Pk_config_cumulant} yields an exact separation
\begin{align}
  \Big\langle e^{\lambda\,\Delta u_z}\,D\,D'\Big\rangle
  =
  W(\lambda,\rv)\;
  \Big\langle e^{\lambda\,\Delta u_z}\,D\,D'\Big\rangle_c\,,
\end{align}
and
\begin{align}
  W(\lambda,\rv)\equiv \Big\langle e^{\lambda\,\Delta u_z}\Big\rangle
  = \exp\!\Big[\big\langle e^{\lambda\,\Delta u_z}\big\rangle_c\Big]\,,
  \label{eq:vdg_def}
\end{align} 
where $\langle\cdots\rangle_c$ denotes a connected correlator. The prefactor $W$ is the \emph{velocity-difference generating function} (VDG): successive derivatives of $\ln W$ with
respect to $\lambda$ generate the cumulants of LOS velocity differences and thus control the non-Gaussian damping
associated with Fingers-of-God. The key modelling choice is to treat $W$ as a prescribed, non-perturbative function rather than expanding it as a power
series in $\lambda$, which would revert to the standard redshift-space PT expansion. 

On large scales, the dominant impact of the VDG is well approximated by its infinite-separation limit, where it depends
only on $\lambda$ and not on $\rv$.  In this approximation the two-point VDG becomes
\begin{equation}
  W(\lambda,\rv)\ \simeq\ W^P(\lambda)
  = \frac{1}{\sqrt{1+\tfrac{1}{2}a_{\rm vir}^2\lambda^2}}\,
  \exp\!\left[
    \frac{\lambda^2\,\sigma_v^2}{1+\tfrac{1}{2}a_{\rm vir}^2\lambda^2}
  \right]\,,
  \label{eq:WP}
\end{equation}
where $a_{\rm vir}$ is a free parameter controlling the kurtosis contribution to the LOS velocity-difference
statistics, and $\sigma_v^2 \equiv \langle u_z^2\rangle$ is the (bulk-flow) velocity dispersion. 
In linear theory, $\sigma_v^2$ is computed from the velocity divergence power spectrum (or equivalently $P_{\rm lin}$)
as 
\begin{equation}
  \sigma_v^2 = \frac{1}{3}\int\frac{\mathrm{d}^3q}{(2\pi)^3}\,\frac{P_{\rm lin}(q)}{q^2}\,.
  \label{eq:sigmav}
\end{equation}

For the bispectrum, an analogous manipulation yields expressions involving a three-point VDG,
$W_{1,2}(\rv_{13},\rv_{23})\equiv\langle e^{\lambda_1 u_{13,z}+\lambda_2 u_{23,z}}\rangle$.
In the infinite-separation approximation we adopt 
\begin{align}
  W_{1,2}(\rv_{13},\rv_{23})\ & \simeq\ W^B(\lambda_1,\lambda_2,\lambda_3)
 \notag \\  &
 =  \frac{1}{\Big(1+\tfrac{1}{2}a_{\rm vir}^2\lambda_{123}^2\Big)^{3/2}}\,
  \exp\!\left[
    \frac{\lambda_{123}^2\,\sigma_v^2}{1+\tfrac{1}{2}a_{\rm vir}^2\lambda_{123}^2}
  \right]\,
\end{align}
where $\lambda_i \equiv i f k_{i,z}$ and $\lambda_{123}^2\equiv \lambda_1^2+\lambda_2^2+\lambda_3^2$ (with
$\kv_1+\kv_2+\kv_3=\mathbf{0}$ for closed triangles).

\subsection{Perturbative expansion}
\label{sec:perturbative_expansion}
Having isolated the VDG contribution, all remaining terms in the cumulant-expanded expressions (i.e.\ the connected
correlators in Eq.~\eqref{eq:vdg_def}) are treated perturbatively. 
Concretely, we model the deterministic clustering contributions using perturbation theory for the matter and velocity
fields, together with an Eulerian galaxy bias expansion.  We adopt the standard EFT bias basis used in \cite{EggemeierEtAl2025}, writing the galaxy overdensity as

\begin{align}
        \delta_{\rm g}(\xv)  = \ & b_1\,\delta(\xv) +\frac{b_2}{2}\delta^{\,2}(\xv) + b_{\nabla^2} \nabla^2 \delta(\xv) 
\nonumber \\
& + b_{\mathcal{G}_2}\mathcal{G}_2\left(\Phi_\mathrm{v}\,|\,\xv\right) + b_{\Gamma_3}\Gamma_3(\xv)\,,
   \label{eq:deltag_expansion}
\end{align}
where  $\mathcal{G}_2$ and 
$\Gamma_3$ are non-local operators, defined as 
\begin{align}    
    &\mathcal{G}_2 (\Phi |\textbf{x}) \vcentcolon = \big [\partial_i \partial_j \Phi(\textbf{x}) \big]^2 - [\partial^2 \Phi(\textbf{x})]^2, \\ \notag \\ 
    &\Gamma_3(\textbf{x}) \vcentcolon = \mathcal{G}_2(\Phi |\textbf{x}) - \mathcal{G}_2(\Phi_v |\textbf{x})\,,
\end{align}
and $\Phi(\textbf{x})$ and $\Phi_v(\textbf{v})$ represent the gravitational and velocity potential.
where $\delta$ is the matter overdensity and $\mathcal{G}_2$ denotes the Galilean-invariant tidal operator, while
$b_{\nabla^2\delta}$ is the leading higher-derivative bias parameter capturing spatial non-locality. 

\subsection{Modelling the redshift-space power spectrum and two-point correlation function}
\label{sec:vdg_pk}
The configuration-space multipoles of the 2PCF are related to the 
power spectrum multipoles via the Hankel transform
\begin{equation}
  \xi_{s,\ell}(r) = i^\ell \int \frac{\mathrm{d}k}{2\pi^2}\,k^2\,
  j_\ell(kr)\,P_{s,\ell}(k)\,.
  \label{eq:xiL_from_PL}
\end{equation}
The VDG model for the redshift-space galaxy power spectrum can be written
schematically as
\begin{align}
  P_s^{\rm VDG}(\kv) = 
  W^P(\kv)\,\Big[P_s^{\rm tree}(\kv) + P_s^{\rm 1\mbox{-}loop}(\kv) 
  + P_s^{\rm ctr}(\kv)\Big]
  + \Delta P(\kv)\,,
  \label{eq:vdg_pk_schematic}
\end{align}
where $W^P$ is the two-point VDG damping function (in the
infinite-separation approximation), and
\begin{align}
  P_s^{\,\rm tree}(\kv) & = Z_1^2(\kv)\,P_{\rm lin}(k)\,, \\
  P_s^{\,\rm 1\mbox{-}loop}(\kv) & = P_{{s},22}(k)+P_{{s},13}(k) \notag \\
    = & \ 2\int_{\qv}Z_2^{\,2}({\qv,\kv-\qv})\,P_{\rm lin}({|\kv-\qv|})\,P_{\rm lin}(q)
    \notag \\
    & + 6\,Z_1(\kv)\,P_{\rm lin}(k)\int_{\qv} Z_3({\qv,-\qv,\kv})\,P_{\rm lin}(q)\,,\\
  P_s^{\rm ctr}(k,\mu) & =
    -\,2\,k^2\,P_{\rm lin}(\kv)\sum_{n=0}^{2} c_{2n}\,\mathcal{L}_{2n}(\mu)\,,\\
  \Delta P(\kv) & =
    -\frac{\lambda^2}{2}\,\sigma_v^2\,P_{DD}(k)
    + \frac{\lambda^2}{2}\int_{\qv}
    \frac{q_z^2}{q^4}\,P_{\theta\theta}(q)\,P_{DD}(|\kv-\qv|)\,.
\end{align}

Stochastic contributions are analytic in $k^2$ in Fourier space: 
a constant shot-noise term maps in configuration space to a Dirac 
delta function $\delta^{(3)}(\mathbf{r})$, terms proportional to 
$k^2$ map to its Laplacian $\nabla^2\delta^{(3)}(\mathbf{r})$, and 
higher-order terms in $k^{2n}$ to its higher spatial derivatives. 
In all cases, these contributions are localised at vanishing separation 
and do not affect the correlation functions at finite separations within 
the EFT regime of validity.

\subsection{Modelling the redshift-space bispectrum and three-point correlation function}
\label{sec:vdg_bk}
The corresponding configuration-space multipoles of the 3PCF are 
related to the bispectrum multipoles via
\begin{align}    
  \zeta_{\ell}(r_{12},r_{13})
  = (-1)^\ell \int \frac{\mathrm{d}k_1\,\mathrm{d}k_2}{(2\pi)^6}\,
  k_1^2\,k_2^2\, B_{\ell}(k_1,k_2)\,
  j_\ell(k_1 r_{12})\,j_\ell(k_2 r_{13})\,,
  \label{eq:zeta_from_B_internal}
\end{align}
which is evaluated efficiently using 2D-FFTLog-based algorithms
\citep[e.g.][]{FangEiflerKrause2020, Umeh2021, GuidiEtal2023, 
FarinaEtal2024a, PugnoEtAl2024a}.
At leading (tree) order in PT, the VDG model for the bispectrum takes the
form \citep{EggemeierEtAl2025}
\begin{align}
  B_s^{\rm VDG}(\kv_1,\kv_2,\kv_3)
  = W^B(\kv_1,\kv_2,\kv_3)\,B^{\rm tree}(\kv_1,\kv_2,\kv_3)\,,
  \label{eq:vdg_bk_schematic}
\end{align}
where $W^B$ is the three-point VDG damping function in the
infinite-separation approximation, and $B^{\rm tree}$ is the standard
SPT tree-level redshift-space galaxy bispectrum
\citep{IvanovSibiryakov2018},
\begin{align}
  B^{\rm tree}(\kv_1,\kv_2,\kv_3)
  = 2\,Z_1(\kv_1)\,Z_1(\kv_2)\,Z_2(\kv_1,\kv_2)\,
  P_{\rm lin}(k_1)\,P_{\rm lin}(k_2) + {\rm cyc.}\,,
  \label{eq:Btree}
\end{align}
where ${\rm cyc.}$ denotes cyclic permutations of $(\kv_1,\kv_2,\kv_3)$. The stochastic contributions to the bispectrum are analytic in $k^2$ 
in Fourier space: a constant shot-noise term maps in configuration space 
to a Dirac delta function localised at coincident points, terms 
proportional to $k^2$ to its Laplacian, and higher-order terms in 
$k^{2n}$ to higher spatial derivatives. In all cases, these contributions 
are non-zero only when two or more vertices of the triangle collapse to 
the same position. Since in practice the 3PCF is measured only at finite, 
non-zero separations, these contributions never enter the analysis and 
can therefore be safely neglected.
\subsection{IR resummation}
\label{sec:IR-resummation}
To account for the smearing of the BAO feature induced by large-scale
displacements, we implement infrared (IR) resummation in the
wiggle/no-wiggle approach \citep{EisensteinSeoWhite2007,
SmithScoccimarroSheth2007, CrocceScoccimarro2008, Matsubara2008A,
DesjacquesEtal2010, BaldaufEtal2015B, SenatoreZaldarriaga2015}.
The linear power spectrum is decomposed as
$P_{\rm lin}(k) = P_{\rm nw}(k) + P_{\rm w}(k)$,
and only the oscillatory component is damped,
\begin{equation}
  P_{\rm lin,IR}(k) = P_{\rm nw}(k) 
  + e^{-k^2\Sigma_{\rm tot}^2}\,P_{\rm w}(k)\,.
  \label{eq:PlinIR}
\end{equation}
In redshift space the displacement variance depends on the line-of-sight
direction and reads
\begin{equation}
  \Sigma_{\rm tot}^2(\mu) =
  \left(1+f\,\mu^2\right)^2\Sigma_1^2
  + f^2\mu^2\!\left(1-\mu^2\right)\Sigma_2^2\,,
  \label{eq:Sigma_tot}
\end{equation}
where $\mu\equiv\hat{\kv}\!\cdot\!\hat{\zv}$ and
\begin{align}
  \Sigma_1^2 &=
    \frac{1}{6\pi^2}\int_0^{k_s}\!dq\,
    \left[1-j_0(q\,\ell_{\rm BAO})+2\,j_2(q\,\ell_{\rm BAO})\right]
    P_{\rm nw}(q)\,,
    \label{eq:Sigma1}\\
  \Sigma_2^2 &=
    \frac{1}{2\pi^2}\int_0^{k_s}\!dq\;
    j_2(q\,\ell_{\rm BAO})\,P_{\rm nw}(q)\,,
    \label{eq:Sigma2}
\end{align}
with $\ell_{\rm BAO}\simeq110\,h^{-1}{\rm Mpc}$ and
$k_s=0.14\,{\rm Mpc}^{-1}$ \citep{EggemeierEtAl2025}.

IR resummation is implemented by replacing $P_{\rm lin}\to P_{\rm lin,IR}$
throughout Sections~\ref{sec:vdg_pk}--\ref{sec:vdg_bk} at three-level. At one-loop
order for the power spectrum, this substitution requires an additional
correction term \citep{IvanovSibiryakov2018},
\begin{equation}
  P_{1\text{-loop}}(k) =
  P^{\rm SPT}_{1\text{-loop}}\!\left[P_{\rm lin,IR}\right](k)
  - k^2\Sigma_{\rm tot}^2\,e^{-k^2\Sigma_{\rm tot}^2}\,
  Z_1^2(\kv)\,P_{\rm w}(k)\,.
  \label{eq:P1loop_IR}
\end{equation}


\section{The BOSS DR12 dataset}
\label{sec:dataset}
Our analysis is based on the final galaxy samples of the Baryon 
Oscillation Spectroscopic Survey (BOSS) Data Release 12 (DR12) 
\footnote{\url{https://www.sdss3.org/index.php}}, part of the 
SDSS-III programme \citep{DawsonEtal2013}. BOSS obtained 
spectroscopy for luminous galaxies over a contiguous footprint 
of $\sim 10^4\,{\rm deg}^2$ split between the North and South 
Galactic Caps (hereafter, NGC and SGC), targeting samples 
designed to trace the large-scale distribution of matter and 
measure the baryon acoustic oscillation (BAO) feature over the 
redshift range $0.15 \lesssim z \lesssim 0.7$.

\subsection{Redshift samples from DR12}

The BOSS DR12 large-scale structure catalogues include four 
target selections: LOWZ, LOWZE2, LOWZE3, and CMASS 
\citep{ReidEtal2012, BeutlerEtal2017}. The LOWZ selection 
targets massive early-type galaxies at low redshift using 
colour--colour and colour--magnitude cuts designed to produce 
an approximately volume-limited sample. The CMASS selection 
targets more distant luminous red galaxies (LRGs) with cuts 
optimised for approximately constant stellar mass. The LOWZE2 
and LOWZE3 selections are extensions of LOWZ designed to 
improve completeness at intermediate redshifts.

Following \citet{ReidEtAl2016}, we use the combined four 
selections into two non-overlapping redshift bins, hereafter 
referred to as \textit{low-z} and \textit{high-z}. The 
combination is performed by resampling the original catalogues 
in redshift, retaining each galaxy in the bin corresponding to 
its spectroscopic redshift. The \textit{low-z} sample covers 
$0.15 < z < 0.43$, with effective redshift 
$z_{\rm eff,\textit{low-z}} \simeq 0.38$, while the 
\textit{high-z} sample spans $0.43 < z < 0.70$, with 
$z_{\rm eff,\textit{high-z}} \simeq 0.61$. In both cases we 
follow the standard BOSS data-quality cuts and veto masks. The 
resulting samples contain of order a few $10^5$ galaxies for 
\textit{low-z} and $\sim 7 \times 10^5$ for \textit{high-z} 
when combining NGC and SGC.

\subsection{Random catalogues and weights}

To characterise the survey geometry and radial selection function we use the public BOSS DR12 random catalogues associated with the \textit{low-z} and \textit{high-z} samples \citep{ReidEtAl2016}. It is a catalog of unclustered objects that follow the same angular mask and redshift distribution as the data, with a density $\sim 50$ times higher than the corresponding galaxy catalogues. They enter all our estimators of two- and three-point statistics. All galaxies are assigned the standard BOSS total weight
\begin{equation}
  w_{\rm tot} = (w_{\rm rf} + w_{\rm fc} - 1)\,w_{\rm sys}\,,
\end{equation}
where $w_{\rm rf}$ and $w_{\rm fc}$ correct for redshift failures and fibre collisions, respectively, and $w_{\rm sys}$ accounts for large-scale angular systematics such as stellar density and seeing variations \citep{RossEtal2012,AndersonEtal2014}. 

\subsection{MultiDark-Patchy mock catalogues}
\label{sec:patchy}

\footnotetext{\url{https://www.skiesanduniverses.org/page/page-3/page-15/page-9/}}

To estimate the covariance matrix of our clustering measurements and to validate
the data analysis pipeline, we make use of the publicly available suite of 2048
MultiDark-Patchy mock catalogues \citep{Klypin2016, KitauraEtal2016},
hereafter referred to as Patchy mocks. These were generated using an approximate
gravity solver calibrated against full N-body simulations, and are designed to
reproduce the statistical properties of the BOSS DR12 galaxy samples and the known selection effects. They are
constructed for all four data chunks (NGC/SGC $\times$ \textit{low-z}/\textit{high-z}), adopting
the same angular mask, veto flags and radial selection function as the BOSS DR12
data described in Sec.~\ref{sec:dataset}. 

Each mock catalogue is accompanied by a dedicated random catalogue sharing the
same geometry as the data. Galaxy weights in the Patchy mocks are assigned as
\begin{equation}
  w_{\rm tot}^{\rm mock} = w_{\rm veto}\,w_{\rm fc}\,,
\end{equation}
comprising a veto mask term and a fibre-collision correction, but omitting the
angular systematic weight $w_{\rm sys}$, which is not required for synthetic
catalogues. 

The procedure adopted to estimate the covariance matrix from the Patchy suite, including the Hartlap correction for the finite number of realisations, is
detailed in Sec.~\ref{sec:covariance}.
\section{Parameter inference}

\label{sec:inference}

We describe here the statistical framework adopted to infer cosmological
and nuisance parameters in a Bayesian approach from the 2PCF and 3PCF measurements presented in
Appendix ~\ref{app:2pcf_measurements} and Appendix \ref{app:3pcf_measurements}. We detail the estimation of the covariance
matrix (Sect.~\ref{sec:covariance}), the data-vector compression scheme
adopted for the joint analysis (Sect.~\ref{sec:cov-compression}), the
likelihood and sampling procedure (Sect.~\ref{sec:likelihood}), and the
goodness-of-fit diagnostics used to validate and compare the different
analysis configurations (Sect.~\ref{sec:fom}).

\subsection{Covariance matrix}
\label{sec:covariance}

We estimate the covariance matrix of the data vector $\mathbf{d}$ numerically,
by applying the same 2PCF and 3PCF estimators described in
App.~\ref{app:estimators} to the full suite of $N_m = 2048$ Patchy mock
catalogues. The sample covariance matrix is estimated as
\begin{equation}
  \hat{C}_{ij} = \frac{1}{N_m - 1}
  \sum_{k=1}^{N_m} \left(d_i^k - \bar{d}_i\right)\!\left(d_j^k - \bar{d}_j\right)\,,
  \label{eq:cov}
\end{equation}
where $d_i^k$ is the value of the $i$-th bin in the $k$-th mock and
$\bar{d}_i$ is the sample mean over all realisations.
The matrix has a natural block structure,
\begin{equation}
  \hat{C} =
  \begin{pmatrix}
    \hat{C}_{\xi\xi} & \hat{C}_{\xi\zeta} \\
    \hat{C}_{\zeta\xi} & \hat{C}_{\zeta\zeta}
  \end{pmatrix},
  \label{eq:cov_blocks}
\end{equation}
where $\hat{C}_{\xi\xi}$ and $\hat{C}_{\zeta\zeta}$ quantify the covariant
errors on the 2PCF and 3PCF multipoles respectively, and
$\hat{C}_{\xi\zeta} = \hat{C}_{\zeta\xi}^T$ encodes their cross-covariance.

Because $N_m$ is finite, the naive inverse $\hat{C}^{-1}$ is a biased estimator
of the true precision matrix. We apply the multiplicative correction of
\citet{HartlapEtal2009},
\begin{equation}
  \hat{\Psi} = \alpha_H \ \hat{C}^{-1},
  \label{eq:hartlap}
\end{equation}
where 
\begin{equation}
    \alpha_H =  \frac{N_m - N_d - 2}{N_m - 1}
\end{equation} 
where $N_d$ is the size of the data vector. Furthermore, the finite number of
mocks introduces an additional uncertainty in $\hat{C}$ that propagates into
the posterior widths of the inferred parameters. Following \citet{PercivalEtal2014},
we correct the marginalised parameter variances by a factor
\begin{equation}
  \alpha_P = \frac{1 + B(N_d - N_p)}{1 + A + B(N_p + 1)}\,,
  \label{eq:percival}
\end{equation}
where $N_p$ is the number of free parameters and
\begin{align}
  A = &\frac{2}{(N_m - N_d - 1)(N_m - N_d - 4)}\,, \qquad \\
  B = &\frac{N_m - N_d - 2}{(N_m - N_d - 1)(N_m - N_d - 4)}\,.
  \label{eq:percival_AB}
\end{align}

\subsection{Optimal data-vector compression}
\label{sec:cov-compression}

For the joint 2PCF+3PCF analysis, the combined data vector has dimension
$N_d$ large enough that the ratio $N_m / N_d$ is insufficient to guarantee
a well-conditioned covariance matrix inversion. Rather than inflating the bin sizes or discarding 3PCF
configurations, we follow \citet{PhilcoxEtal2021} and compress the data vector
into a low-dimensional subspace that retains, by construction, the vast
majority of the cosmological information. The method proceeds by constructing a
\emph{template bank} of model vectors, analysing their covariance structure
with Singular Value Decomposition (SVD), and retaining only the modes that
carry the largest prior-averaged cosmological information. This yields an
ordered basis in data space, where the retained modes span the directions
most relevant to the inference and the discarded ones contribute
negligibly to the likelihood.

The \emph{template bank} consists of $N_{\rm bank} =
10^4$ model evaluations drawn uniformly from the prior on the full cosmological
and nuisance parameter space $\boldsymbol{\theta} = \{10^9A_s,\, h,\,
\omega_{\rm cdm},\, b_1,\, b_2,\, b_{G_2},\, b_{\Gamma_3},\, c_0,\, c_2,\,
c_4,\, a_{\rm vir}\}$. Each model vector $\mathbf{d}(\boldsymbol{\theta}^{(i)})$
is evaluated using the theory emulators described in Sec.~\ref{sec:emulation}, making the generation of the full bank
computationally feasible. The template spectra are first noise-weighted and
mean-subtracted,
\begin{equation}
  X_{\alpha}^{(i)} = \sum_{b} C^{-1/2}_{\alpha b}
  \left[d_b(\boldsymbol{\theta}^{(i)}) - \bar{d}_b\right],
  \label{eq:svd_rotation}
\end{equation}
where $\bar{\mathbf{d}}$ is the mean over all templates and $C^{1/2}$ is  the Cholesky decomposition of the sample covariance. Stacking the $N_{\rm bank}$ rotated vectors into a matrix $\mathbf{X}$
of dimension $N_{\rm bank} \times N_d$ and performing the Singular Value Decomposition (SVD),
$\mathbf{X} = \mathbf{U}\,\mathbf{D}\,\mathbf{V}^T$, yields an ordered
set of basis vectors $\{\mathbf{V}_\alpha\}$ and singular values
$\{D_\alpha\}$. The basis vectors are ordered by decreasing singular
value, so that the first modes span the directions in data space that
carry the most prior-averaged cosmological information; modes with small
$D_\alpha$ contribute negligibly to the likelihood and can be safely
discarded.

The number of retained modes $N_{\rm SV}^*$ is determined by the criterion
\begin{equation}
  N_{\rm SV}^* = \min \left( N_{\rm SV} \;\bigg|\;
  \frac{1}{N_{\rm bank} \cdot N_{\rm dof}(N_{\rm SV})}
  \sum_{\alpha > N_{\rm SV}} D_\alpha^2
  \leq \Delta\tilde\chi^2_{\rm red,\,thr} \right),
  \label{eq:svd_threshold}
\end{equation}
with threshold $\Delta\tilde\chi^2_{\rm red,\,thr} = 0.05$, where
$N_{\rm dof}(N_{\rm SV})$ is the number of degrees of freedom corresponding
to the retained subspace. This is a slight variation of the criterion explained in
\citet{PhilcoxEtal2021}, where we normalise by the number of degrees of
freedom so that the threshold is expressed in terms of the mean
\emph{reduced} $\tilde\chi^2$ error incurred by the subspace projection,
averaged across the prior. 

The compressed covariance $\hat{C}^{\rm sub}_{\alpha\beta}$ is estimated
directly from the 2048 Patchy mocks in the projected space, and the
Hartlap and Percival corrections of Eqs.~\eqref{eq:hartlap}--\eqref{eq:percival}
are applied with $N_d$ replaced by $N_{\rm SV}^*$, yielding corrections
significantly closer to unity than in the uncompressed case.

\subsection{Likelihood and MCMC sampling}
\label{sec:likelihood}
We perform the cosmological inference in a Bayesian framework, sampling
the posterior distribution of the model parameters $\boldsymbol{\theta}$.
The posterior is proportional to the product of the likelihood and the
prior,
\begin{equation}
  P(\boldsymbol{\theta}|\mathbf{d}) \propto \mathcal{L}(\mathbf{d}|\boldsymbol{\theta})\,
  \pi(\boldsymbol{\theta})\,,
\end{equation}
where $\pi(\boldsymbol{\theta})$ denotes the prior distribution. We adopt a Gaussian likelihood for the data vector $\mathbf{d}$,
\begin{equation}
  -2\ln\mathcal{L}(\boldsymbol\theta)
  = \chi^2(\boldsymbol\theta)
  = \big(\mathbf{d} - \mathbf{m}(\boldsymbol\theta)\big)^T\,
    \hat{\Psi}\,
    \big(\mathbf{d} - \mathbf{m}(\boldsymbol\theta)\big)\,,
  \label{eq:likelihood}
\end{equation}
For the 2PCF-only analysis, $\mathbf{d} =
\boldsymbol{\xi}$ and $\hat{\Psi} = \hat{\Psi}_{\xi\xi}$. For the joint
2PCF+3PCF analysis, the data vector is the concatenation
$\mathbf{d} = (\boldsymbol{\xi},\,\boldsymbol{\zeta})^T$ and the precision
matrix is computed from the full joint covariance of Eq.~\eqref{eq:cov_blocks},
which retains the cross-covariance block $\hat{C}_{\xi\zeta}$; in this case,
following the procedure described in Sec.~\ref{sec:cov-compression}, $\hat\Psi$ is
obtained from the Hartlap-corrected inverse of the compressed covariance
$\hat{C}^{\rm sub}$.

We sample the posterior distribution using the \textsc{emcee} code
\citep{ForemanMackeyEtal2013}, which implements an affine-invariant ensemble
sampler. We employ $N_{\rm walk}$ walkers and terminate the chains only
after each walker has accumulated at least 100 integrated autocorrelation
times $\tau_{\rm ac}$, ensuring full convergence. Different priors are assumed for all free parameters; the complete
list of parameters and prior ranges is given in Tab.~\ref{tab:priors}.

\subsection{Emulation of full-shape redshift-space templates}
\label{sec:emulation}

The affine-invariant ensemble sampler described above requires 
$\mathcal{O}(10^5\text{--}10^6)$ evaluations of the theoretical 
data vector during a typical MCMC run. Direct evaluation of the 
configuration-space templates at every step is computationally 
prohibitive: while the 2PCF multipoles can be computed via 
one-dimensional Hankel transforms in a fraction of a second, the 
3PCF requires a two-dimensional FFTLog integration over a large 
number of triangle configurations, making a brute-force approach 
unfeasible for likelihood sampling purposes. A fast emulator is 
therefore an essential ingredient of the analysis, and what makes 
a full-shape configuration-space joint analysis of the 2PCF and 
3PCF practically achievable.

To this end, we construct a neural-network emulator that predicts, 
as a function of the cosmological and nuisance parameters 
$\boldsymbol{\theta}$, the full joint data vector
\begin{equation}
  \mathbf{d} \;=\;
  \Big\{
    \xi_{s,0}(r),\;\xi_{s,2}(r),\;\xi_{s,4}(r),\;
    \zeta_{s,0,\ell}(r_{12},r_{13})
  \Big\}_{\ell=0}^{\ell_{\max}},
\end{equation}
comprising the monopole, quadrupole and hexadecapole of the 2PCF 
and the $(\ell_{\max}+1)$ Legendre multipoles of the isotropic 3PCF 
evaluated over all admissible triangle configurations $(r_{12},r_{13})$. The emulator architecture follows \citet{EuclidGuidiEtAl2025}, to 
which we refer the reader for full implementation details. With 
respect to that work, the framework is extended to account for 
redshift-space distortions and the geometric Alcock--Paczy\'nski 
distortions, parametrised by the dilation parameters
\begin{equation}
  \alpha_\perp
    = \frac{D_A(\boldsymbol{\theta})\,r_s^{\rm fid}}
           {D_A^{\rm fid}\,r_s(\boldsymbol{\theta})},
  \qquad
  \alpha_\parallel
    = \frac{H^{\rm fid}\,r_s^{\rm fid}}
           {H(\boldsymbol{\theta})\,r_s(\boldsymbol{\theta})}.
\end{equation}
where $D_A(z)$ and $H(z)$ are the angular diameter distance,  
Hubble parameter at the effective redshift of each sample, $r_s$ is the sound horizon 
at the drag epoch and the fiducial cosmology has been specified in Tab \ref{tab:patchy_cosmo}. The training set is constructed via a Latin Hypercube Sampling (LHS) 
scheme over the prior volume in $(h,\,\omega_{\rm cdm},\,10^9A_s,\,a_{\rm vir}) \in [(0.65, 0.75), (0.08, 0.015), (1.5, 2.6), (0, 15)]$, 
with $10^4$ training points for the 2PCF and $4\times10^3$ for the 3PCF. 
The smaller training set for the 3PCF is justified by the smoother 
parameter dependence of the 3PCF templates over the explored prior 
volume, as verified by the accuracy diagnostics described below.

\begin{figure}[htbp]
\centering
\begin{subfigure}{0.49\columnwidth}
    \includegraphics[width=\linewidth]{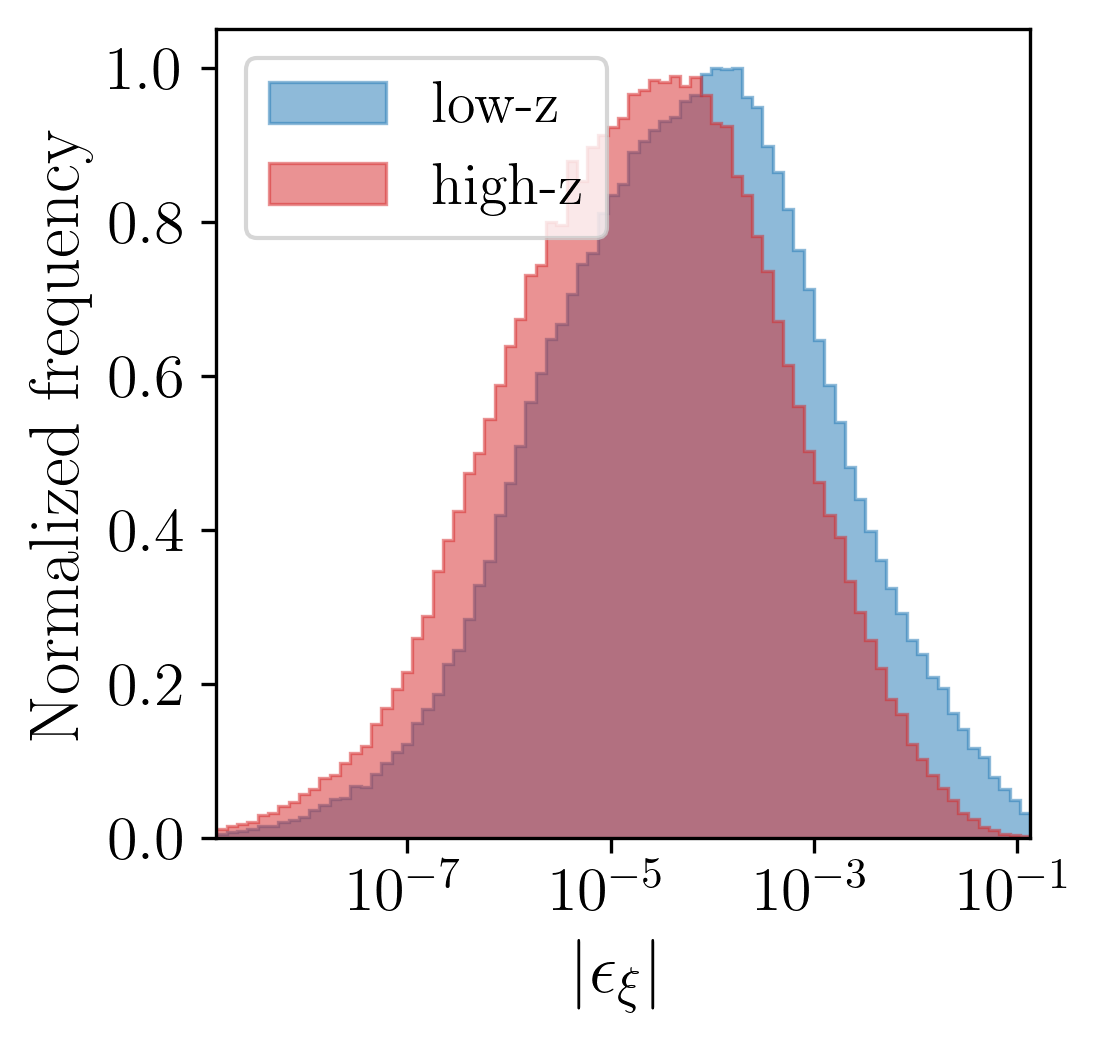}
    \caption{2PCF multipoles.}
    \label{fig:emu_acc_xi}
\end{subfigure}
\hfill
\begin{subfigure}{0.475\columnwidth}
    \includegraphics[width=\linewidth]{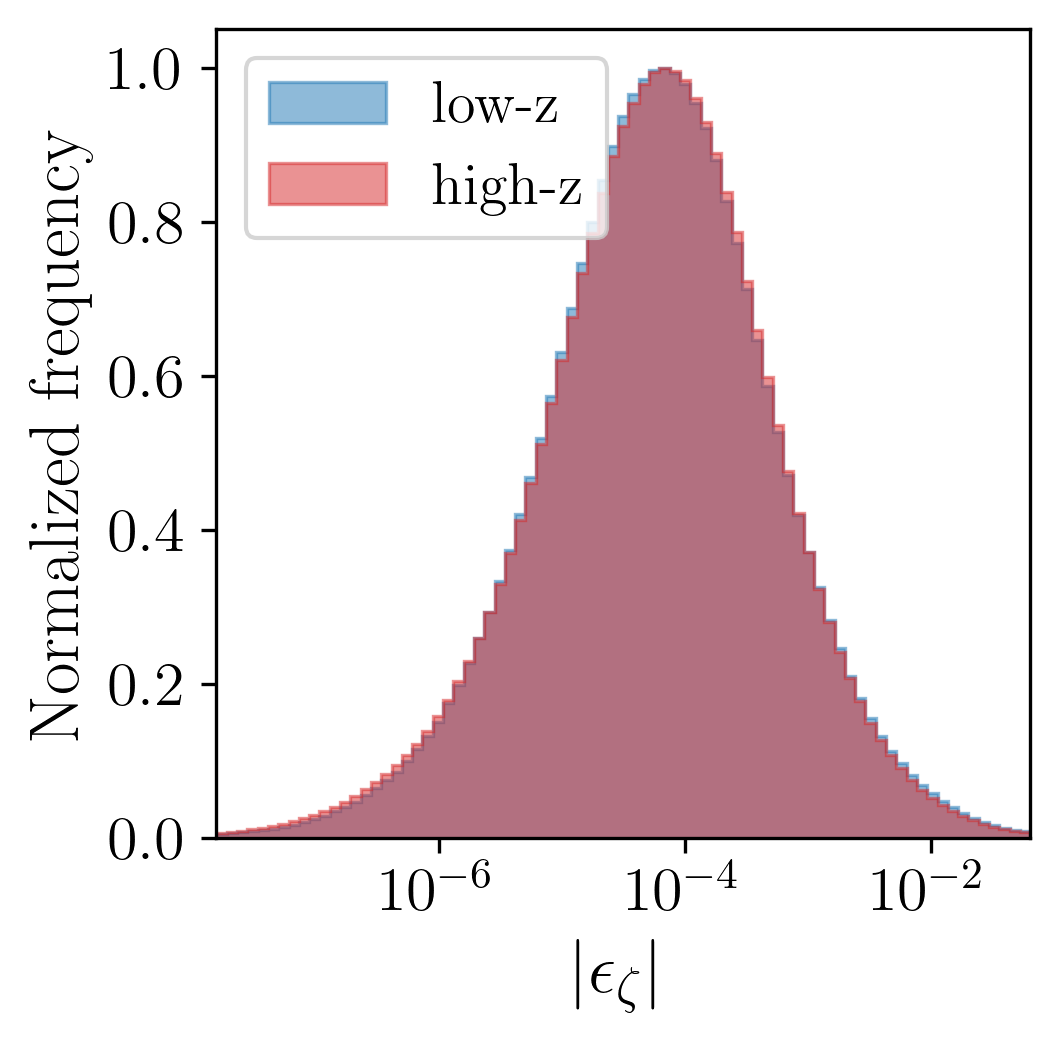}
    \caption{3PCF multipoles.}
    \label{fig:emu_acc_zeta}
\end{subfigure}
\caption{Distribution of normalised emulator residuals 
$|\epsilon|$ for the 2PCF multipoles ($\ell = 0, 2, 4$; left) 
and 3PCF multipoles ($\ell = 0, 1, 2, 3, 4$; right), evaluated 
over a test set for the \textit{low-z} (blue) and 
\textit{high-z} (red) BOSS DR12 samples. The residuals are 
normalised by the diagonal of the data covariance matrix, 
$\sigma_\xi(r)$ and $\sigma_\zeta(r_{12},r_{13})$, and thus 
quantify the emulator systematic as a fraction of the statistical 
error.}
\label{fig:emu_acc_combined}

\end{figure}

The emulator accuracy is assessed on a test set not used 
during training. For each test point, we evaluate the normalised 
residual of every element of the data vector,
\begin{align}
  \epsilon_\xi(r) \;\equiv\; &
    \frac{\xi^{\rm emu}(r) - \xi^{\rm exact}(r)}{\sigma_\xi(r)},
  \qquad \\
  \epsilon_\zeta(r_{12}, r_{13}) \;\equiv\; &
    \frac{\zeta^{\rm emu}(r_{12},r_{13})
          - \zeta^{\rm exact}(r_{12},r_{13})}
         {\sigma_\zeta(r_{12},r_{13})},
\end{align}
which directly quantifies the fraction of the statistical error 
budget consumed by the emulator systematic. The normalised distributions of 
$|\epsilon_\xi|$ and $|\epsilon_\zeta|$ across all test points, 
separation bins, and multipoles are shown in 
Figs.~\ref{fig:emu_acc_xi} and~\ref{fig:emu_acc_zeta} for the 
\textit{low-z} and \textit{high-z} samples. In all cases the 
distributions peak at $|\epsilon| \sim 10^{-4}\text{--}10^{-5}$, 
with tails not exceeding $|\epsilon| \lesssim 10^{-1}$, implying 
that the emulator systematic is smaller than the statistical 
uncertainty by at least three to four orders of magnitude at the 
median, and remains below $10\%$ of $\sigma$ everywhere in the 
explored parameter space. A mild improvement in accuracy is observed 
for the \textit{high-z} sample in the 2PCF residuals, likely 
reflecting the different scale ranges and signal-to-noise ratios of 
the two samples, while the 3PCF residuals are comparable between the 
two bins. We therefore conclude that the emulator introduces no 
appreciable systematic bias into the likelihood analysis.

\subsection{Goodness of fit and Figure of Merit}
\label{sec:fom}
To assess the goodness of fit of each analysis configuration, we compute 
the reduced chi-squared
\begin{equation}
  \chi^2_{\rm red} = \frac{\chi^2(\hat{\boldsymbol\theta})}{N_{\rm dof}}\,,
  \label{eq:chi2red}
\end{equation}
where $\hat{\boldsymbol\theta}$ denotes the maximum a posteriori (MAP) 
estimate, i.e.\ the point in the parameter space that maximises the posterior 
$p(\boldsymbol\theta\,|\,\boldsymbol{d})$, and $\chi^2(\hat{\boldsymbol\theta})$ 
is evaluated at that point. The number of degrees of freedom is 
$N_{\rm dof} = N^*_{\rm SV} - N_p$, where $N_p$ is the number of free 
parameters. A value of $\chi^2_{\rm red}$ consistent with unity within 
the expected statistical fluctuations of order $\sqrt{2/N_{\rm dof}}$ 
indicates that the model provides an adequate description of the data; 
systematic deviations above or below this range signal model breakdown 
or data over-fitting, respectively.

To quantify and compare the constraining power of the different analyses
(2PCF-only and joint 2PCF$+$3PCF), we adopt the Figure of Merit
\citep{AlbrechtEtal2006a}
\begin{equation}
  {\rm FoM}(\boldsymbol\theta_{\rm cos})
  = \frac{1}{\sqrt{\det\!\left[S(\boldsymbol\theta_{\rm cos})\right]}}\,,
  \label{eq:fom}
\end{equation}
where $S(\boldsymbol\theta_{\rm cos})$ is the parameter covariance matrix
of the cosmological subset $\boldsymbol\theta_{\rm cos} = \{10^9A_s,\,
h,\, \omega_{\rm cdm}\}$, marginalised over all nuisance parameters and
extracted from the MCMC chains. The FoM is inversely proportional to the
volume of the confidence hyper-ellipsoid in cosmological parameter space:
a larger FoM corresponds to tighter, less degenerate constraints. We
evaluate both $\chi^2_{\rm red}$ and the FoM as a function of the minimum
scale cut and discuss the results in Sec.~\ref{sec:2pcf_analysis} and Sec. \ref{sec:joint_analysis}.


\begin{table}
  \centering
  \footnotesize
  \caption{Prior distributions for all free parameters. $\mathcal{U}[a,b]$
  denotes a uniform prior; $\mathcal{N}(\mu,\sigma^2, a,b)$ a Gaussian
  truncated to $[a,b]$. Cosmological parameters are shared between redshift
  bins in the \textit{combined samples} analysis; all nuisance parameters are
  independent per bin.}
  \label{tab:priors}
  \begin{tabular}{llc}
    \hline\hline
    Parameter & Description & Prior \\
    \hline
    \multicolumn{3}{c}{\textit{Cosmological parameters}} \\
    \hline
    $10^9 A_s$         & Scalar amplitude           & $\mathcal{N}(2.145,\,0.5^2, 1.0,3.0)$  \\
    $h$                & Hubble parameter           & $\mathcal{U}[0.65,\,0.75]$                  \\
    $\omega_{\rm cdm}$ & CDM density                & $\mathcal{U}[0.09,\,0.15]$                  \\
    \hline
    \multicolumn{3}{c}{\textit{Galaxy bias parameters}} \\
    \hline
    $b_1$              & Linear bias                & $\mathcal{U}[1.0,\,3.0]$                    \\
    $b_2$              & Quadratic bias             & $\mathcal{N}(0,\,1^2, -5,5)$            \\
    $b_{G_2}$          & Non-local bias             & $\mathcal{N}(0,\,1^2, -5,5)$            \\
    $b_{\Gamma_3}$     & Third-order non-local bias & $\mathcal{N}(0.548,\,1^2, -5,5)$        \\
    \hline
    \multicolumn{3}{c}{\textit{EFT counter-terms}} \\
    \hline
    $c_0$              & Isotropic counter-term     & $\mathcal{N}(0,\,10^2, -10^4,10^4)$     \\
    $c_2$              & Quadrupole counter-term    & $\mathcal{N}(0,\,10^2, -10^3,10^3)$     \\
    $c_4$              & Hexadecapole counter-term  & $\mathcal{N}(0,\,10^2, -10^3,10^3)$     \\
    \hline
    \multicolumn{3}{c}{\textit{Velocity dispersion}} \\
    \hline
    $a_{\rm vir}$      & Virial velocity parameter  & $\mathcal{N}(0,\,20^2, 0,10^3)$         \\
    \hline\hline
  \end{tabular}
\end{table}

\section{Cosmological constraints}
\label{sec:2pcf_analysis}

The 2PCF and 3PCF multipoles measured from the BOSS DR12 data are presented
in Appendix~\ref{app:measurements}; in this section we show the
cosmological inference analysis.

\subsection{2PCF}
\subsubsection{Analysis setup}
\label{sec:2pcf_setup}

We perform a full-shape likelihood analysis of the 2PCF multipoles measurements described
in Appendix \ref{app:2pcf_measurements}, fitting simultaneously the monopole,
quadrupole and hexadecapole ($\ell = 0, 2, 4$) using the model described in Sec. \ref{sec:theory_model} and the inference framework of Sec.~\ref{sec:inference}.
The free parameters are the three cosmological parameters
$\{10^9 A_s,\, h,\, \omega_{\rm cdm}\}$ and the nuisance parameters
$\{b_1,\, b_2,\, b_{G_2},\, b_{\Gamma_3},\, c_0,\, c_2,\, c_4,\, a_{\rm vir}\}$,
all sampled with priors as listed in Tab.~\ref{tab:priors}. 
All remaining cosmological parameters, including the baryon density $\omega_b \equiv \Omega_b h^2$, are held fixed at the fiducial values of
Tab.~\ref{tab:patchy_cosmo}. Following similar analyses, the priors adopted in this analysis are intentionally broad and are chosen primarily to ensure numerical stability and to restrict the sampler to physically plausible regions of parameter space. In particular, the cosmological priors on $10^9A_s$, $h$, and $\omega_{\rm cdm}$ are wide enough that the posterior constraints are data-driven, while the nuisance priors reflect theoretical expectations from perturbation theory and calibration from simulations. The baryon density $\omega_b$ is fixed to the fiducial Patchy value for internal consistency with the mock-based forward model and to avoid introducing an additional degeneracy in the 2PCF-only fit.
The analysis is performed separately for the \textit{low-z} and \textit{high-z}
redshift bins, combining the NGC and SGC sub-samples within each bin under the
assumption of shared bias parameters. We also perform a joint
\textit{low-z}+\textit{highz-z} analysis,  hereafter \textit{combined samples}, in
which the two redshift bins share the cosmological parameters but have
independent sets of nuisance parameters.
To assess the sensitivity of the results to the choice of minimum scale,
we repeat every analysis for $r_{\rm min} \in \{20, 30, 40, 50, 60, 70, 80\}
\,h^{-1}{\rm Mpc}$ keeping the maximum scale fixed at $r_{\rm max} = 130\,h^{-1}{\rm Mpc}$.

\subsubsection{Cosmological parameter constraints}
\label{sec:2pcf_cosmo}

\begin{figure}[ht]
  \centering
  \includegraphics[width=1\columnwidth]{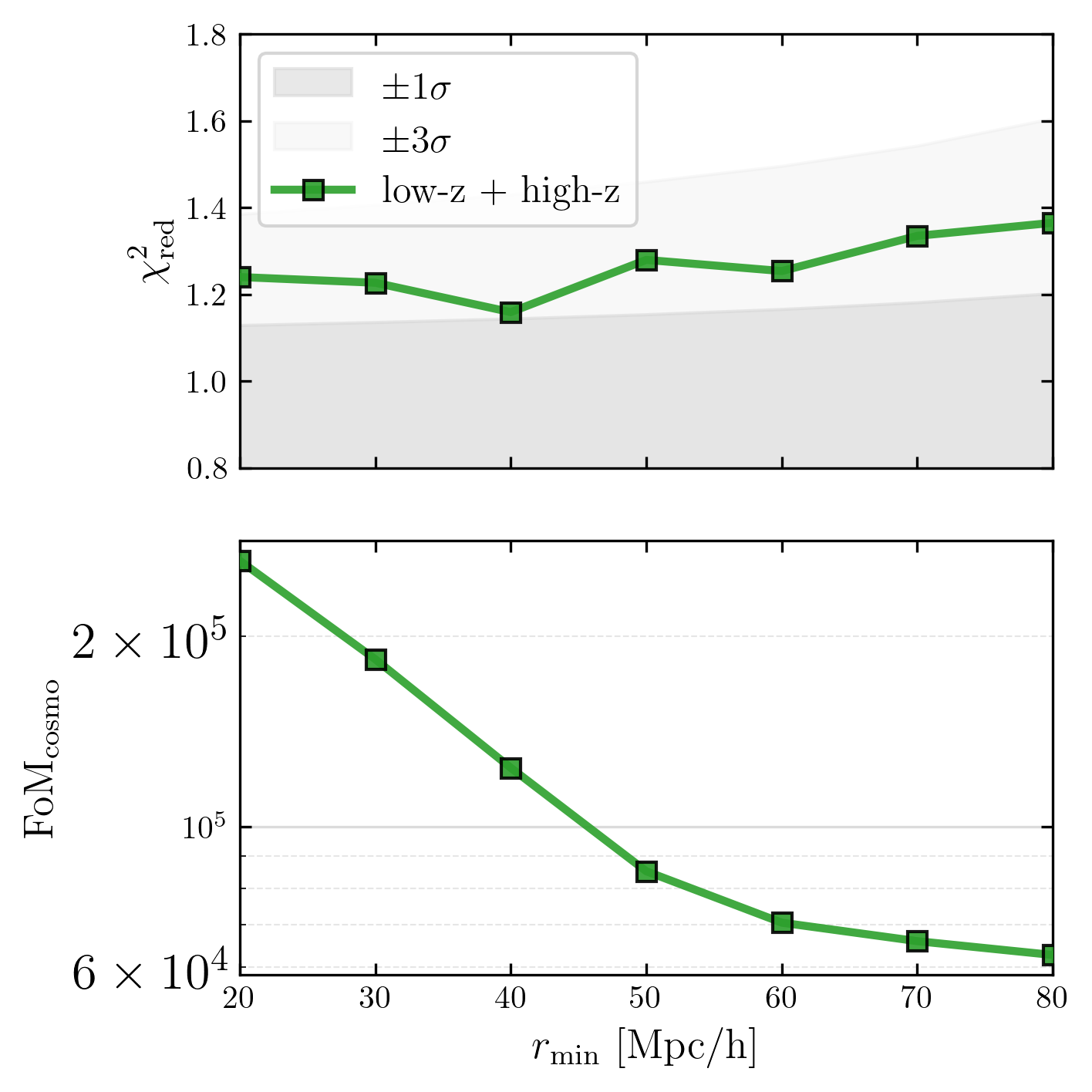}
  \caption{Model diagnostics for the \textit{combined samples} 2PCF analysis as a
function of $r_{\rm min}$, with $r_{\rm max} = 130\,h^{-1}{\rm Mpc}$ fixed.
\textit{Top panel:} reduced chi-squared $\chi^2_{\rm red}$ at the posterior
mean; dark and light grey bands mark the $\pm1\sigma$ and $\pm3\sigma$
expected uncertainties. \textit{Bottom panel:} Figure of Merit on the cosmological subset 
$\{10^9 A_s,\, h,\, \omega_{\rm cdm}\}$.}
  \label{fig:chi2_vs_rmin_2PCF}
\end{figure}

We begin by assessing the validity range of the perturbative model through
the diagnostics shown in Fig.~\ref{fig:chi2_vs_rmin_2PCF}, which presents
the reduced chi-squared $\chi^2_{\rm red}$ and the figure of merit on the
cosmological subspace, hereafter FoM$_{\rm{cosmo}}$, as a function of
$r_{\rm min}$ for the \textit{combined samples} analysis. The data-vector
sizes and the corresponding Hartlap and Percival correction factors for all
scale cuts are reported in Tab.~\ref{tab:2pcf_corrections} of
App.~\ref{app:scale_cut_2pcf}.

The upper panel shows that $\chi^2_{\rm red}$, evaluated at the posterior
mean, remains comfortably within the $\pm 2\sigma$ band expected for the
corresponding number of degrees of freedom across the entire range of
$r_{\rm min}$ explored. While this confirms that the model provides an
acceptable fit at all scale cuts, the goodness-of-fit criterion alone does
not uniquely identify the regime of validity of the perturbative model.
We therefore complement this analysis with a study of the stability of the
marginalised posteriors as a function of $r_{\rm min}$, presented in
App.~\ref{app:scale_cut_2pcf}. Based on the $\chi^2$ goodness-of-fit
criterion, we adopt $r_{\rm min} = 20\,h^{-1}{\rm Mpc}$ as the reference
scale cut for all subsequent 2PCF analyses, in analogy with previous
analyses of the same dataset \citep{SanchezEtal2017a}.

\begin{figure}[ht]
  \centering
  \includegraphics[width=\columnwidth]{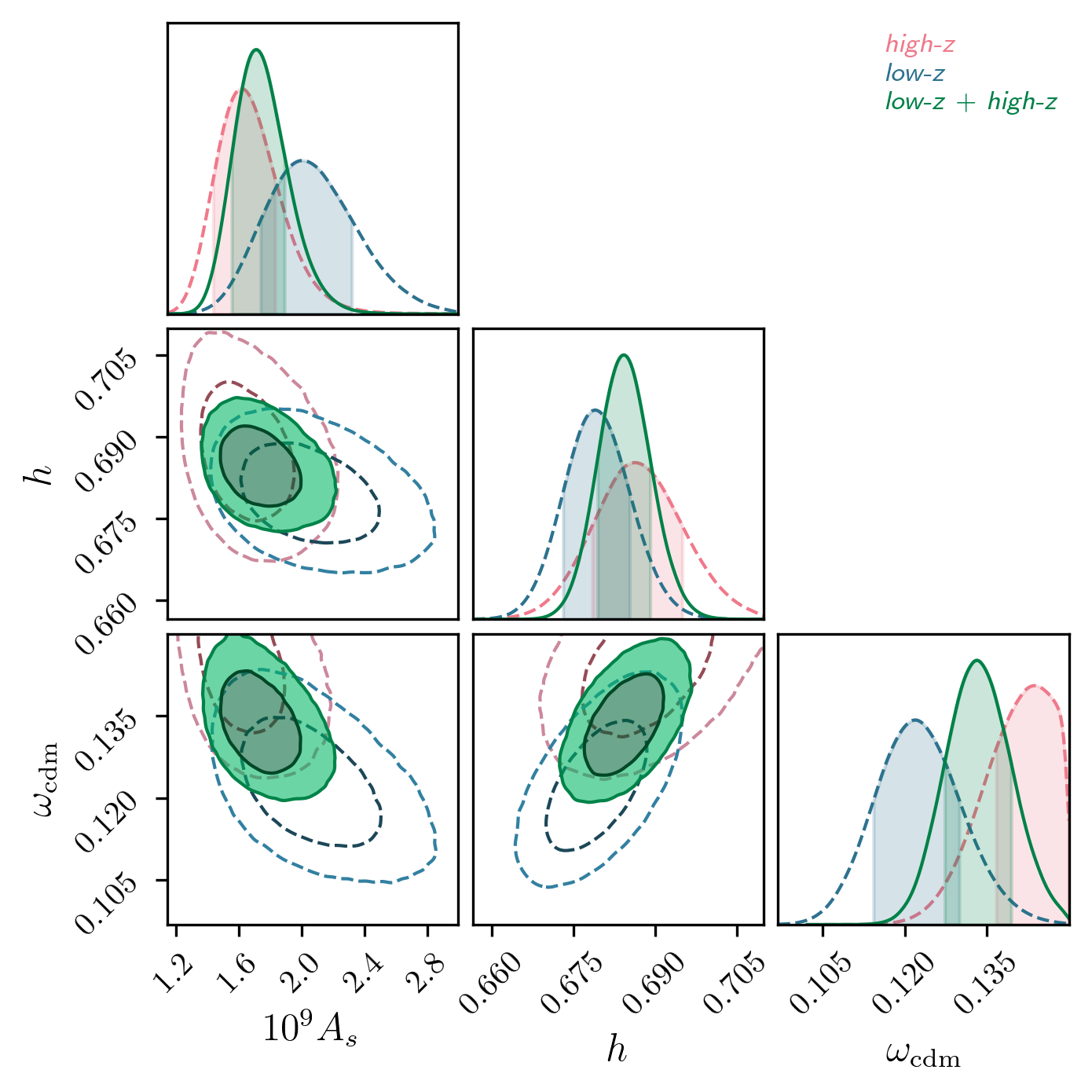}
  \caption{Marginalised posterior distributions of the cosmological parameters
  $10^9 A_s$, $h$ and $\omega_{\rm cdm}$ from the 2PCF-only analysis at the
  reference scale cut $r_{\rm min} = 20\,h^{-1}{\rm Mpc}$. The diagonal
  panels show the one-dimensional marginal distributions; the off-diagonal
  panels show the joint 68\% and 95\% credible regions. Filled green contours
  correspond to the \textit{combined samples} redshift bins analysis; dashed blue and
  red contours show the individual \textit{low-z} and \textit{high-z} results,
  respectively.}
  \label{fig:chains_cosmo_2pcf}
\end{figure}

The lower panel shows FoM$_{\rm{cosmo}}$, which quantifies the
constraining power of the analysis in the $\{10^9 A_s,\, h,\,
\omega_{\rm cdm}\}$ space. It decreases steeply and monotonically as
$r_{\rm min}$ increases, reflecting the progressive loss of information as
the data vector shrinks. In particular, the FoM$_{\rm{cosmo}}$ at
$r_{\rm min} = 20\,h^{-1}{\rm Mpc}$ is approximately three times larger than
that at $r_{\rm min} = 40\,h^{-1}{\rm Mpc}$, highlighting the significant
gain in cosmological information retained when smaller scales are included in
the analysis.

The marginalised posteriors of the cosmological parameters at
$r_{\rm min} = 20\,h^{-1}{\rm Mpc}$ are shown in
Fig.~\ref{fig:chains_cosmo_2pcf}. The diagonal panels display the
one-dimensional marginal distributions of $10^9 A_s$, $h$ and
$\omega_{\rm cdm}$, while the off-diagonal panels show the joint $68\%$ and
$95\%$ credible contours for the \textit{low-z} (dashed blue),
\textit{high-z} (dashed red) and \textit{combined samples} (filled green)
analyses. The \textit{low-z} and \textit{high-z} bins are analysed separately and are 
broadly consistent within statistical uncertainties. The \textit{high-z} 
posterior for $\omega_{\rm cdm}$ is slightly broader than the \textit{low-z} 
one, with the marginalised distribution extending toward the upper boundary of 
the prior range; this behaviour is a known feature of full-shape analyses of 
single BOSS DR12 redshift bins, and has been observed in analogous analyses 
\citep{IvanovEtal2022B}. It reflects the limited constraining power of the 
high-z sample alone on the broad-band shape of the 2PCF, which encodes most 
of the $\omega_{\rm cdm}$ information, rather than a systematic bias in the 
modelling pipeline. The \textit{combined samples} analysis exploits the 
complementarity between the two redshift bins to break this degeneracy, 
yielding substantially tighter and well-converged constraints:
$10^9 A_s = 1.736^{+0.183}_{-0.155}$,
$h = 0.684 \pm 0.005$ and
$\omega_{\rm cdm} = 0.133^{+0.006}_{-0.006}$.

The Hubble parameter is the best-constrained of the three, with a
${\sim}\,0.7\%$ precision, while $10^9 A_s$ carries the largest relative
uncertainty. The 2D contours reveal a positive $h$--$\omega_{\rm cdm}$
correlation, driven by the joint sensitivity of the BAO scale, and a
negative $A_s$--$\omega_{\rm cdm}$ correlation. The latter, discussed further
in App.~\ref{app:scale_cut_2pcf}, underlies the observed shift of
$\omega_{\rm cdm}$ as smaller scales are included in the analysis. As
broad-band shape information is progressively added to the data vector, the
degeneracy between these parameters evolves, leading to a corresponding
broadening and shift of their marginalised posteriors.

\subsection{Joint 2PCF + 3PCF}
\label{sec:joint_analysis}
We present here the full-shape cosmological analysis combining the 2PCF and
3PCF multipoles of the BOSS DR12 \textit{combined samples} data, whose
measurements are shown in Appendix~\ref{app:measurements}. The 2PCF scale
cut is fixed at $r_{\rm min}^{\rm 2PCF} = 20\,h^{-1}{\rm Mpc}$ as
established in Sect.~\ref{sec:2pcf_analysis}. For the 3PCF we explore a
range of $r_{\rm min}^{\rm 3PCF} \in [40, 80]\,h^{-1}{\rm Mpc}$ and
$\eta_{\rm min} \in \{1, 2, 3\}$, where $\eta_{\rm min}$ is a threshold
on the elongation of the retained triangle configurations, defined as
\begin{equation}
    \eta \equiv \frac{r_{13} - r_{12}}{\Delta r}\,,
    \label{eq:eta}
\end{equation} 
with $\Delta r$ being the separation bin size, to parametrise the proximity of a given configuration to the isosceles case. Fixing a lower limit, $\eta_{\rm min}$, amounts to excluding isosceles, or nearly isosceles configurations \citep{VeropalumboEtal2022}. In particular, the case
$\eta_{\rm min} = 0$ is excluded, as it retains nearly equilateral
triangle configurations that are problematic for perturbative inference
at tree level; further details are given in App.~\ref{app:estimators}.

\begin{figure}
  \centering
  \includegraphics[width=\columnwidth]{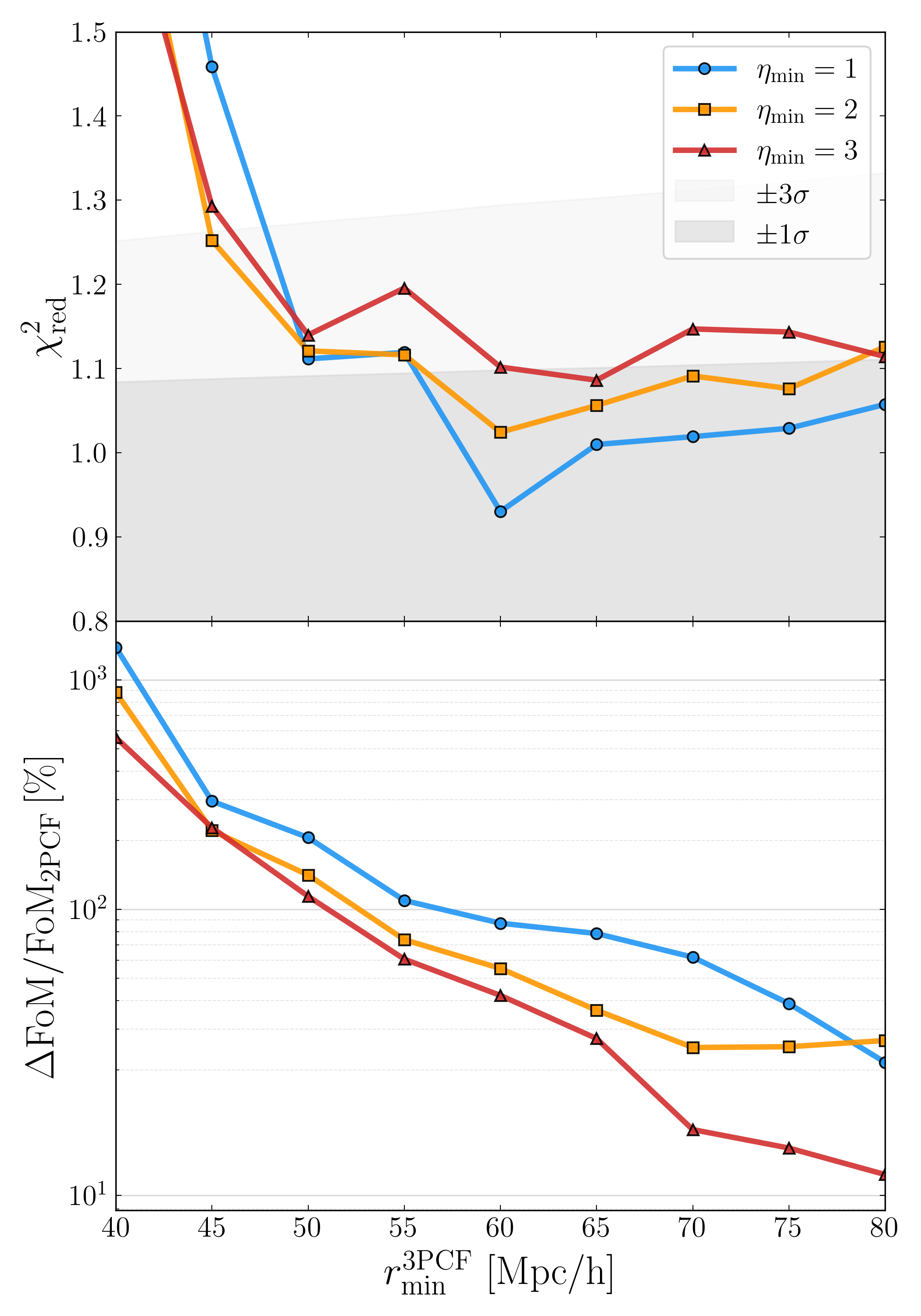}
  \caption{Goodness-of-fit and FoM gain of the joint 2PCF+3PCF
  \textit{combined samples} analysis as a function of $r_{\rm min}^{\rm 3PCF}$,
  with $r_{\rm min}^{\rm 2PCF} = 30\,h^{-1}{\rm Mpc}$ fixed. Results are
  shown for $\eta_{\rm min} = 1$ (blue), $2$ (green), and $3$ (orange).
  \textit{Upper panel:} reduced chi-squared $\chi^2_{\rm red}$ evaluated
  at the posterior mean; grey bands mark the $\pm1\sigma$ and $\pm3\sigma$
  expected intervals for the corresponding number of degrees of freedom, at the fixed value $\eta_{\rm min} = 3$.
  \textit{Lower panel:} relative FoM gain $\Delta{\rm FoM}/{\rm FoM}_{\rm
  2PCF}$ in percent on a logarithmic scale, where ${\rm FoM}_{\rm 2PCF}$
  is the figure of merit of the 2PCF-only analysis at  $r_{\rm min}^{\rm
  2PCF} = 20\,h^{-1}{\rm Mpc}$.}
  \label{fig:chi2_fomgain}
\end{figure}

\subsubsection{Scale cut selection}
\label{sec:joint_scalecut}

Figure~\ref{fig:chi2_fomgain} summarises the two diagnostics used to
identify the reference scale cut for the joint analysis. The upper panel
shows the reduced chi-squared $\chi^2_{\rm red}$, evaluated at the
posterior mean, as a function of $r_{\rm min}^{\rm 3PCF}$ for the three
values of $\eta_{\rm min}$. The lower panel shows the relative FoM gain
$\Delta{\rm FoM}/{\rm FoM}_{\rm 2PCF}$ in percent, where
${\rm FoM}_{\rm 2PCF}$ represents the figure of merit of the 2PCF-only
analysis at $r_{\rm min}^{\rm 2PCF} = 20\,h^{-1}{\rm Mpc}$.

At $r_{\rm min}^{\rm 3PCF} = 40\,h^{-1}{\rm Mpc}$ the $\chi^2_{\rm red}$
exceeds the $\pm3\sigma$ expected band for all three $\eta_{\rm min}$
values. It decreases steeply as $r_{\rm min}^{\rm 3PCF}$ increases;
$\eta_{\rm min}=1$ and $\eta_{\rm min}=2$ enter the $\pm1\sigma$ band
around $r_{\rm min}^{\rm 3PCF}\simeq55$-$60\,h^{-1}{\rm Mpc}$, while
$\eta_{\rm min}=3$ remains slightly above the $\pm1\sigma$ band across
most of the explored range, reflecting the fact that fewer triangle
configurations leave less freedom to absorb residual model imperfections.
The goodness of fit is therefore driven primarily by $r_{\rm min}^{\rm
3PCF}$, with a secondary dependence on $\eta_{\rm min}$ that is most
visible at intermediate scale cuts. We note that the expected bands
themselves are largely insensitive to $\eta_{\rm min}$, as the number
of degrees of freedom varies negligibly across the three choices.

The lower panel reveals a complementary picture. The FoM gain decreases
monotonically as $r_{\rm min}^{\rm 3PCF}$ increases, and a clear
hierarchy among the three elongation cuts is present across the entire
range: $\eta_{\rm min}=1 > \eta_{\rm min}=2 > \eta_{\rm min}=3$. This
ordering directly reflects the information content of the different
triangle subsets: smaller $\eta_{\rm min}$ admits a larger number of
configurations, in particular squeezed and nearly-degenerate triangles,
which carry additional cosmological signal. At $r_{\rm min}^{\rm
3PCF}=60\,h^{-1}{\rm Mpc}$ the FoM gains are approximately $90\%$,
$55\%$ and $40\%$ for $\eta_{\rm min}=1$, $2$ and $3$ respectively;
at $r_{\rm min}^{\rm 3PCF}=70\,h^{-1}{\rm Mpc}$ they reduce to
approximately $60\%$, $25\%$ and $15\%$.

To report our results, we adopt
$r_{\rm min}^{\rm 3PCF}=60\,h^{-1}{\rm Mpc}$ with $\eta_{\rm min}=1$
as a reference configuration for the joint analysis: this is the
a conservative choice on the smallest scale cut at which the model is statistically acceptable, and
it retains the largest FoM gain consistent with perturbative validity.
The validity of this choice is further supported by the stability of the
marginalised posteriors as a function of $r_{\rm min}^{\rm 3PCF}$,
discussed in Sect.~\ref{sec:joint_cosmo}. The configurations
$r_{\rm min}^{\rm 3PCF}=60\,h^{-1}{\rm Mpc}$ with $\eta_{\rm min}=2$
and $3$, and $r_{\rm min}^{\rm 3PCF}=70\,h^{-1}{\rm Mpc}$ with
$\eta_{\rm min}=3$, are used as robustness checks.

\begin{figure}
  \centering
  \includegraphics[width=\columnwidth]{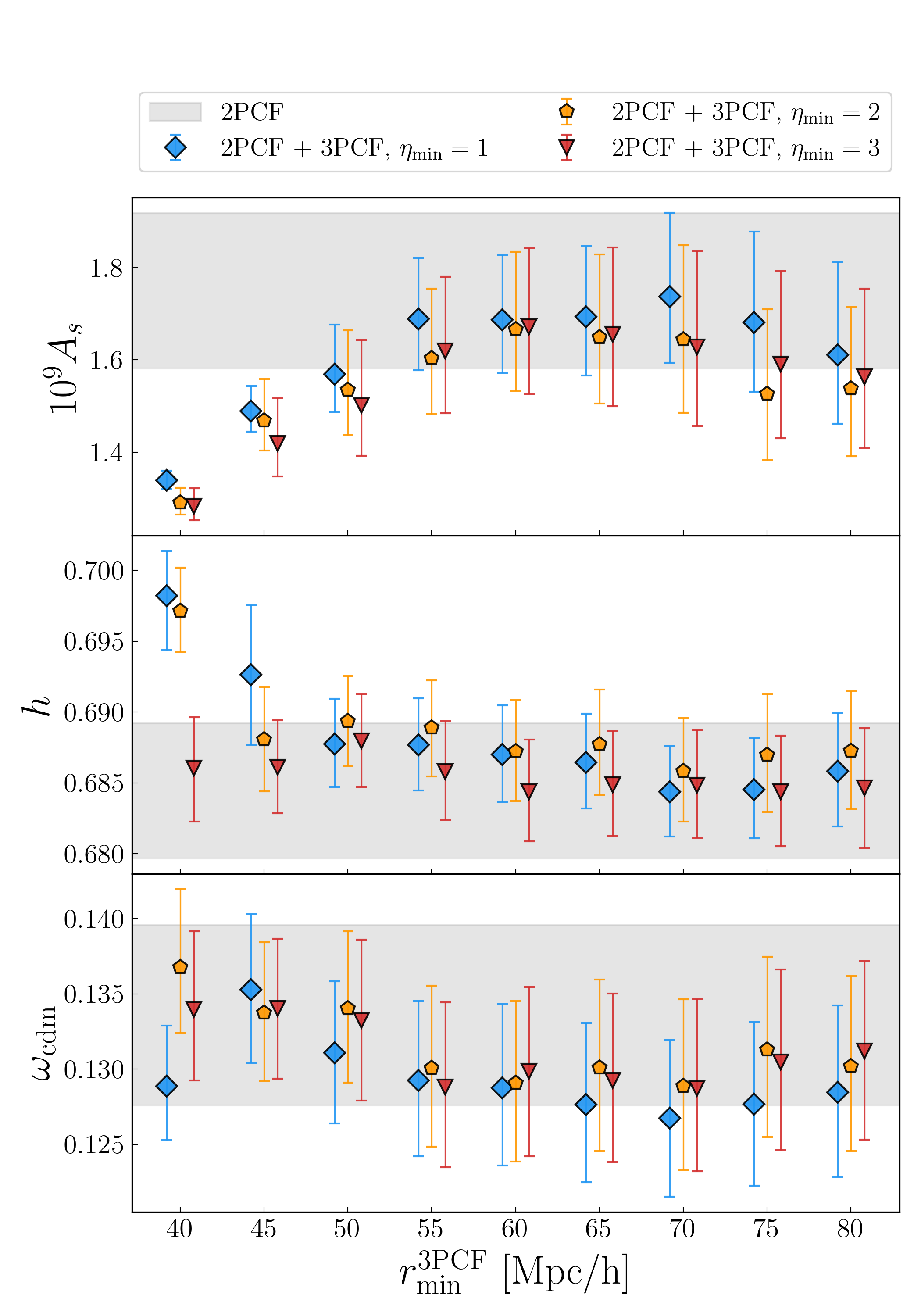}
  \caption{Marginalised posterior means and $68\%$ credible intervals for
  $10^9 A_s$ (top), $h$ (middle) and $\omega_{\rm cdm}$ (bottom) from the
  joint 2PCF+3PCF \textit{combined samples} analysis as a function of $r_{\rm
  min}^{\rm 3PCF}$, for $\eta_{\rm min} = 1$ (blue diamonds), $2$ (green
  pentagons) and $3$ (orange triangles). The grey dashed line and shaded
  band show the posterior mean and $1\sigma$ interval of the 2PCF-only
  reference analysis at $r_{\rm min}^{\rm 2PCF} = 20\,h^{-1}{\rm Mpc}$.
  The dotted line marks the Patchy fiducial value.}
  \label{fig:cosmo_vs_rmin3pcf_joint}
\end{figure}

\begin{figure}
  \includegraphics[width=1\columnwidth]{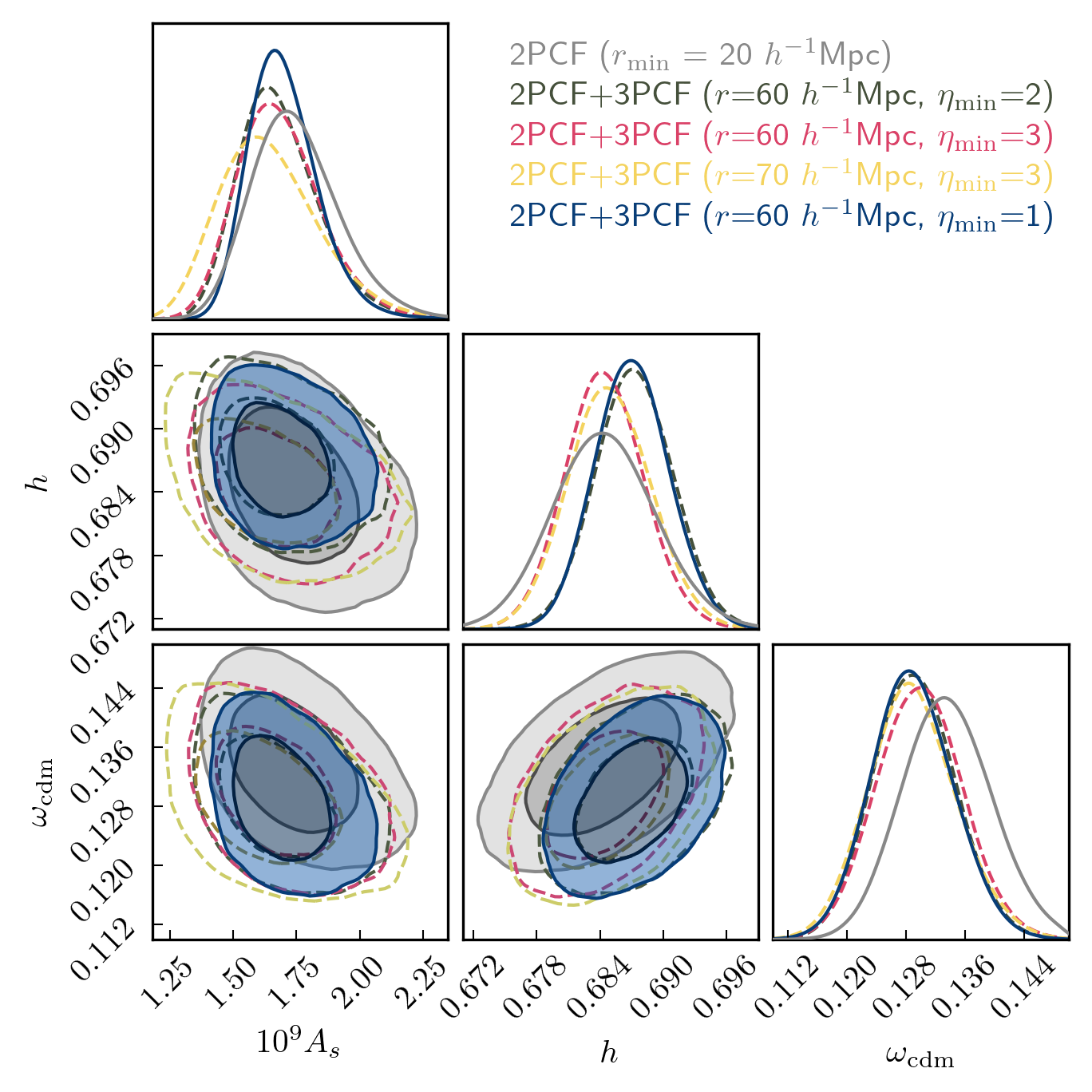}
  \caption{Marginalised posterior distributions of $10^9 A_s$, $h$ and
  $\omega_{\rm cdm}$ for the joint 2PCF+3PCF \textit{combined samples} analysis.
  Grey: 2PCF-only result at $r_{\rm min}^{\rm 2PCF} = 20\,h^{-1}{\rm
  Mpc}$. Filled dark green: reference joint analysis ($r_{\rm min}^{\rm
  3PCF} = 60\,h^{-1}{\rm Mpc}$, $\eta_{\rm min} = 3$). Dashed pink:
  $r_{\rm min}^{\rm 3PCF} = 70\,h^{-1}{\rm Mpc}$, $\eta_{\rm min} = 2$.
  Dashed blue: $r_{\rm min}^{\rm 3PCF} = 70\,h^{-1}{\rm Mpc}$,
  $\eta_{\rm min} = 3$. Different contours enclose the $68\%$ and $95\%$
  credible regions.}
  \label{fig:chains_joint_cosmo}
\end{figure}

\subsubsection{Cosmological constraints}
\label{sec:joint_cosmo}

Figure~\ref{fig:cosmo_vs_rmin3pcf_joint} shows the marginalised posterior
means and $68\%$ and $99.7\%$ credible intervals of the cosmological subset $10^9A_s$, $h$, and
$\omega_{\rm cdm}$ as a function of $r_{\rm min}^{\rm 3PCF}$, for three
choices of the elongation cut $\eta_{\rm min}\in\{1,2,3\}$; the grey
band marks the 2PCF-only reference constraint. At small $r_{\rm min}^{\rm 3PCF}$ values, the three parameters exhibit
a coherent drift with respect to the 2PCF-only baseline: $10^9A_s$
decreases, $h$ increases, and $\omega_{\rm cdm}$ shifts more mildly in
the same direction. This pattern is consistent with the elevated
$\chi^2_{\rm red}$ at those scales (Fig.~\ref{fig:chi2_fomgain}) and
indicates that the joint model is operating close to the edge of its
validated perturbative range. As $r_{\rm min}^{\rm 3PCF}$ increases,
the posterior means converge progressively toward the 2PCF-only band,
and from $r_{\rm min}^{\rm 3PCF}\simeq60\,h^{-1}{\rm Mpc}$ onward all
three parameters are stable and mutually consistent across all
elongation cuts.

Figure~\ref{fig:chains_joint_cosmo} compares the full marginalised
posterior distributions of $\{10^9A_s,\,h,\,\omega_{\rm cdm}\}$ for the
2PCF-only baseline (grey) and for four joint configurations:
$r_{\rm min}^{\rm 3PCF}=60\,h^{-1}{\rm Mpc}$ with $\eta_{\rm min}=1$,
$2$, and $3$ (dark blue, dashed pink, and dashed magenta, respectively),
together with the additional robustness run at
$r_{\rm min}^{\rm 3PCF}=70\,h^{-1}{\rm Mpc}$, $\eta_{\rm min}=3$
(dashed yellow). All considered joint configurations yield mutually consistent
posteriors, confirming the robustness of the results against the specific
choice of $r_{\rm min}^{\rm 3PCF}$ and $\eta_{\rm min}$ within the
validated range.

The inclusion of the 3PCF improves the constraints on all three
cosmological parameters, with a gain that depends on both the scale cut
and the elongation threshold. At the reference configuration
($r_{\rm min}^{\rm 3PCF}=60\,h^{-1}{\rm Mpc}$, $\eta_{\rm min}=1$),
the joint analysis yields
$10^9A_s = 1.687^{+0.141}_{-0.154}$,
$h = 0.687 \pm 0.0035$, and
$\omega_{\rm cdm} = 0.1288^{+0.0056}_{-0.0052}$.
The $1\sigma$ uncertainty on $h$ shrinks from $\pm0.0048$ in the
2PCF-only baseline to $\pm0.0035$, a reduction of $29\%$, making it the
parameter that benefits most from the addition of the 3PCF. The
constraint on $\omega_{\rm cdm}$ improves by $10\%$, while the
$10^9A_s$ posterior narrows by $24\%$, though its central value shifts
by less than $0.3\sigma$ with respect to the 2PCF-only result. This
hierarchy confirms that, at the adopted scale cuts, the 3PCF contributes
primarily through the geometric and redshift-space distortion sector of
parameter space, with an appreciable but secondary effect on the
clustering amplitude.

The gain on $h$ is robust across all configurations explored: even the
most conservative choice ($r_{\rm min}^{\rm 3PCF}=70\,h^{-1}{\rm Mpc}$,
$\eta_{\rm min}=3$) yields a $20\%$ improvement on $\sigma(h)$. The
improvement on $10^9A_s$ is instead more sensitive to the scale cut:
it reaches $24\%$ at $r_{\rm min}^{\rm 3PCF}=60\,h^{-1}{\rm Mpc}$,
but becomes negligible or even slightly negative at larger $r_{\rm
min}^{\rm 3PCF}$, indicating that the amplitude information is
concentrated in the smaller-scale triangle configurations. The central
values of all three parameters are stable across configurations, with
shifts well within the respective $1\sigma$ intervals, confirming the
robustness of the cosmological inference within the validated range.


\section{Discussion and conclusion}
\label{sec:conclusions}
 
We have presented the first full-shape redshift-space joint analysis 
combining the 2PCF and 3PCF on real data, applied to the BOSS DR12 
combined sample across two redshift bins. This analysis constitutes 
a milestone on several fronts:
\begin{itemize}
    \item for the first time, a full-shape cosmological inference 
    jointly combining the 2PCF and 3PCF has been performed on 
    observational data, accounting for both redshift-space distortions 
    and Alcock--Paczy\'nski effects developing a dedicated emulator. This extends to the full 
    redshift-space case the joint analysis previously validated in 
    real space by \citet{EuclidGuidiEtAl2025};

    \item the redshift-space 3PCF is modelled here for the first time 
    within the VDG framework, extending to configuration space the 
    approach developed for the bispectrum in \citet{EggemeierEtAl2025};

    \item optimal data-vector compression \citep{PhilcoxEtal2021} is 
    applied for the first time to a joint 2PCF and 3PCF likelihood 
    analysis, enabling a stable inversion of the joint covariance 
    matrix without discarding any cosmologically informative 3PCF 
    configurations.
\end{itemize}

\subsection{Main results}
 
First, we have performed a full-shape 2PCF-only analysis of the combined
sample. The reference scale cut $r_{\rm min} = 20\,h^{-1}{\rm Mpc}$
was selected through a goodness-of-fit criterion: the reduced
$\chi^2_{\rm red}$ remains statistically acceptable down to this scale,
indicating that the perturbative model provides a good description
of the data, while the posteriors have fully stabilised. The resulting combined-sample constraints are $10^9 A_s = 1.736^{+0.183}_{-0.155}$,
$h = 0.684 \pm 0.005$ and
$\omega_{\rm cdm} = 0.133^{+0.006}_{-0.006}$, 
with the two redshift bins mutually consistent within $1\sigma$ for all
three parameters. The reduced Hubble parameter \textit{h} is the best-constrained of the
three, with a $\sim 0.7\%$ precision already at the two-point level,
driven by the sensitivity of the 2PCF to the BAO scale.
 
For the joint 2PCF and 3PCF analysis, with the 2PCF scale cut fixed at
$r_{\rm min}^{\rm 2PCF} = 20\,h^{-1}{\rm Mpc}$, the reference 3PCF
scale cut was identified by inspecting the goodness-of-fit, considered here ad a performance metric of the modelling as well as the figure of merit, as detailed in Sect.~\ref{sec:joint_scalecut}. The 
diagnostic consistently point to $r_{\rm min}^{\rm 3PCF} =
60\,h^{-1}{\rm Mpc}$ with $\eta_{\rm min} = 1$ as the reference
configuration, which represent our conservative choice on the minimum scale for which the model is
statistically acceptable, and retains a FoM gain of $\sim 90\%$ over the
2PCF-only baseline. The inclusion of the 3PCF improves the constraints on all three
cosmological parameters at aforementioned reference configuration
yielding $10^9A_s = 1.687^{+0.141}_{-0.154}$,
$h = 0.687 \pm 0.0035$, and
$\omega_{\rm cdm} = 0.1288^{+0.0056}_{-0.0052}$.
The largest gain is on $h$, whose uncertainty shrinks by $29\%$,
followed by $A_s$ ($24\%$) and $\omega_{\rm cdm}$ ($10\%$). This
hierarchy reflects the fact that, at the adopted scale cuts, the 3PCF
primarily breaks degeneracies through the angular scale of the BAO
feature, constraining $h$, and through the matter--radiation equality
scale encoded in the shape of the correlation function, constraining
$\omega_{\rm cdm}$, while the contribution along the $A_s$ direction
is limited because the $b_1$--$A_s$ degeneracy is only partially lifted
by the triangle configurations included in the analysis. The gain on $h$
and $\omega_{\rm cdm}$ is stable across all triangle configurations at $r_{\rm
min}^{\rm 3PCF} = 60\,h^{-1}{\rm Mpc}$, while the improvement on
$A_s$ is slightly more sensitive to $\eta_{\rm min}$. Full
details are given in Sect.~\ref{sec:joint_cosmo}.

\subsection{Comparison with the literature}

The results above are fully consistent with the broader picture
emerging from full-shape analyses of BOSS DR12 in Fourier space.
A joint analysis of the power spectrum and bispectrum monopole
of the same dataset was first performed by \citet{GilMarinEtal2015,
GilMarinEtal2017} within a perturbative RSD model, finding a significant
improvement on $f\sigma_8$ from the inclusion of the bispectrum. The
present work extends this analysis to configuration space within a more
complete theoretical framework accounting for both RSD and the AP effect,
inferring the primary cosmological parameters directly in a full-shape
analysis for the first time.

The interpretation requires care given the different parameter spaces.
Fourier-space EFTofLSS analyses typically vary all five primary
$\Lambda$CDM parameters with priors on $\omega_b$ and $n_s$
\citep{IvanovEtal2022B, DAmicoEtal2020, PhilcoxIvanov2022, DAmicoEtal2024},
whereas in the present analysis both are held fixed to the reference
Planck cosmology. Fixing $n_s$ removes the degeneracy with the broad-band
shape, tightening the posteriors on $\omega_{\rm cdm}$ and $A_s$ relative
to the free-$n_s$ case. More importantly for the comparison of $h$
constraints, fixing $\omega_b$ eliminates the degeneracy between the
baryon density and the sound horizon scale, which in analyses that vary
$\omega_b$, even with a BBN or Planck prior, propagates directly into
the $h$ posterior through the BAO measurement. This is the primary reason
why our $\sigma(h)$ is tighter than that reported by Fourier-space analyses
varying the full parameter set, and should be borne in mind when
interpreting the comparison below.

Our constraint on the Hubble parameter, $h = 0.681 \pm 0.005$ from the
2PCF-only analysis and $h = 0.687 \pm 0.003$ from the joint 2PCF$+$3PCF
analysis, is in good agreement with the range $h \simeq 0.68$--$0.70$
consistently reported by Fourier-space full-shape analyses of the same
dataset. The precision achieved is competitive with Fourier-space analyses
that include the bispectrum: \citet{PhilcoxIvanov2022} report $\sigma(h)
\simeq 0.009$ with a Planck prior on $n_s$, while \citet{DAmicoEtal2024}
obtain $\sigma(h) = 0.011$ from the joint one-loop power spectrum and
bispectrum. The inferred cold dark matter density,
$\omega_{\rm cdm} = 0.127 \pm 0.007$ from the 2PCF-only analysis and
$\omega_{\rm cdm} = 0.1288 \pm 0.0054$ from the joint analysis, is in
good agreement with Fourier-space analyses employing a Planck prior on
$n_s$: \citet{PhilcoxIvanov2022} find $\omega_{\rm cdm} = 0.1227 \pm
0.0056$ and \citet{DAmicoEtal2024} obtain $\Omega_m = 0.311 \pm 0.010$
(corresponding to $\omega_{\rm cdm} \approx 0.122$). Without a Planck
prior on $n_s$, Fourier-space analyses find systematically higher values
$\omega_{\rm cdm} \simeq 0.137$--$0.141$ with larger uncertainties
\citep{PhilcoxIvanov2022}, as expected from the $\omega_{\rm cdm}$--$n_s$
degeneracy in the broad-band shape. Our constraint on $A_s$ is broadly
consistent with the amplitude constraints from the same dataset
\citep{PhilcoxIvanov2022, DAmicoEtal2024}, with residual uncertainty
reflecting the $b_1$--$A_s$ degeneracy only partially broken by the
multipole structure of the two- and three-point statistics.

The gain achieved by combining 2PCF and 3PCF is quantitatively consistent
with \citet{SugiyamaEtal2021}, who report a $\sim$30\% reduction in
$\sigma(H(z))$ from the inclusion of the anisotropic 3PCF over the
2PCF pre-reconstruction baseline, decreasing to $\sim$20\% on top of a
reconstructed field. Our analysis recovers a 29\% reduction in $\sigma(h)$,
24\% in $\sigma(A_s)$, and 10\% in $\sigma(\omega_{\rm cdm})$ without
any density-field reconstruction, in close agreement with these predictions.
A direct quantitative comparison with the 15--20\% gain reported in
\citet{AlamEtal2017} requires care, as that result applies to $H(z)\,r_s$
rather than to absolute cosmological parameters; the comparison is therefore
best understood as analogical. Beyond this analogy, the two strategies
differ fundamentally in scope: post-reconstruction BAO sharpens the acoustic
peak but contributes no additional information on $A_s$ or
$\omega_{\rm cdm}$ beyond what the full-shape 2PCF already provides,
whereas the 3PCF simultaneously contributes acoustic scale and broad-band
information. The partial overlap between the two strategies, reflected in
the reduction from $\sim$30\% to $\sim$20\% in \citet{SugiyamaEtal2021}
when the 3PCF is added on top of a reconstructed 2PCF, suggests that
combining 3PCF full-shape measurements with post-reconstruction BAO could
yield further improvements beyond either strategy alone.

\subsection{Future perspectives}
  
This work opens several directions for future investigation. A natural
extension is the inclusion of additional cosmological parameters in the
inference, both within $\Lambda$CDM and beyond, possibily including primordial
non-Gaussianity, for which the three-point statistics are
a primary observable carrying sensitivity to scale-dependent bias on
large scales, and massive neutrinos, whose suppression of small-scale
power leaves correlated imprints on both the 2PCF and the 3PCF. The relevance of higher-order statistics for beyond-$\Lambda$CDM in a similar inference analysis has been further demonstrated by \citet{LuetAl2025}, who find a
preference for evolving dark energy in a joint
one-loop power spectrum and bispectrum analysis of the same dataset explored in this work. This
motivates extending the present framework to beyond-$\Lambda$CDM cosmologies, where the joint 2PCF and 3PCF analysis provides a
natural configuration-space counterpart.A first
step in this direction, exploring neutrino imprints in the 3PCF has been shown in \citet{LabateEtAl2026}. Moreover, the
inclusion of the anisotropic 3PCF with resepct to the line-of-sight would encode additional information on RSD
beyond the monopole retained here, potentially tightening cosmolocial constraints. On the modelling side, extending the one-loop prediction for
the 3PCF from the dark matter case, as shown in \cite{GuidiEtal2023}, to the full
galaxy bias expansion would allow smaller scale cuts to be adopted,
increasing the information content of the analysis and potentially improving the constraining power. On the other hand, the inference approach
developed in this work is directly applicable to next-generation spectroscopic
surveys such as DESI \citep{AdameEtal2024DESI} and \textit{Euclid}
\citep{MellierEtal2024}, where the larger survey volume and
higher galaxy number density will substantially amplify the constraining
power of the joint analysis. 

\begin{acknowledgements}

We thank Azadeh Moradinezhad Dizgah, Farshad Kamalinejad, Zachary Slepian, Kristers Nagainis, and Enzo Branchini for useful discussions. MG and MM acknowledge support from the MIUR PRIN 2022 grant "Optimizing the extraction of cosmological information from Large Scale Structure analysis in view of the next large spectroscopic surveys" (grant $2022NY2ZRS001$). Computational resources were provided by INFN Sezione di Genova, Bologna Physics and Astrophysics Department and Leonardo supercomputing facilities at CINECA through the INAF grants INA24C3B12 and INA24C7B06. We thank the INFN IT personnel in Genova for their continuous support. This project has received funding from the European Union Horizon Europe Research and Innovation Action under grant agreement no. 101183153-WST. Views and opinions expressed are however those of the author(s) only and do not necessarily reflect those of the European Union or the European Research Executive Agency (REA). Neither the European Union nor the REA can be held responsible for them.
\end{acknowledgements}

\bibliographystyle{aa}
\bibliography{cosmology}


\begin{appendix}

\section{Perturbation Theory}
\label{app:decomposition_kernels}

\subsection{Definitions and conventions}
\label{app:definitions}

We define the loop integration measure as
\begin{equation}
  \int_{\qv} \equiv \int \frac{\mathrm{d}^3 q}{(2\pi)^3}\,,
\end{equation}
and adopt the Fourier convention
\begin{align}
  \delta(\kv) &\equiv (2\pi)^3 \int_{\xv} \delta(\xv)\,e^{i\kv\cdot\xv}\,, \\
  \delta(\xv) &\equiv \int_{\kv} \delta(\kv)\,e^{-i\kv\cdot\xv}\,.
\end{align}
Under statistical homogeneity and isotropy in the plane-parallel
approximation, the redshift-space power spectrum is defined by
\begin{equation}
  \langle \delta_s(\kv_1)\,\delta_s(\kv_2)\rangle
  \equiv (2\pi)^3\,\delta_\mathrm{D}(\kv_1+\kv_2)\,P_s(\kv_1)\,,
\end{equation}
and decomposed into Legendre multipoles as
\begin{equation}
  P_{s,\ell}(k) \equiv \frac{2\ell+1}{2}\int_{-1}^{1}\mathrm{d}\mu\,
  P_s(k,\mu)\,\mathcal{L}_\ell(\mu)\,,
  \qquad \mu \equiv \hat{\kv}\cdot\hat{\nv}\,,
\end{equation}
with $\ell = 0, 2, 4$. The redshift-space bispectrum is defined by
\begin{equation}
  \langle \delta_s(\kv_1)\delta_s(\kv_2)\delta_s(\kv_3)\rangle
  \equiv (2\pi)^3\,\delta_\mathrm{D}(\kv_1+\kv_2+\kv_3)\,
  B_s(\kv_1,\kv_2)\,,
\end{equation}
and the redshift-space 3PCF by
\begin{equation}
  \zeta_s(\rv_{12},\rv_{13}) \equiv
  \langle \delta_s(\rv_1)\delta_s(\rv_2)\delta_s(\rv_3)\rangle\,,
\end{equation}
where $\rv_{ij} \equiv \rv_i - \rv_j$. Both statistics depend on two 
distinct angular structures: the orientation of the triangle with respect 
to the line of sight, and the internal shape of the triangle.

\subsection{Angular decomposition of bispectrum and 3PCF}
\label{app:angular_decomposition}

In this work we restrict to the \emph{monopole} of the 3PCF and 
bispectrum over the line-of-sight (LOS) orientation 
\citep{Scoccimarro1997, SlepianEisenstein2015B, SugiyamaEtal2019}, 
i.e.\ we average over all orientations of the triangle with respect 
to the LOS, retaining only the dependence on the internal triangle 
shape. The residual dependence on the internal angle 
$\mu_{12} \equiv \hat{\kv}_1\cdot\hat{\kv}_2$ (or 
$\mu_{12,13} \equiv \hat{\rv}_{12}\cdot\hat{\rv}_{13}$ in 
configuration space) is then expanded in Legendre polynomials up to 
a given $\ell_{\max}$.

\paragraph{Bispectrum.}
Following \citet{Scoccimarro1997}, the bispectrum is first expanded in
spherical harmonics with respect to the orientation of $\kv_1$ relative
to the LOS $\nv$,
\begin{equation}
  B_s(\kv_1,\kv_2) = \sum_{L,M} B_{s,L}^M(k_1,k_2,\mu_{12})\,
  Y_L^M(\omega_{k_1},\theta_{k_1})\,,
\end{equation}
where $\omega_{k_1}$ and $\theta_{k_1}$ are the polar and azimuthal angles
of $\kv_1$ with respect to $\nv$, and we have assumed invariance
under rotations in the plane orthogonal to $\nv$. The multipole
coefficients are
\begin{equation}
  B_{s,L}^M(k_1,k_2,\mu_{12}) \equiv \int \mathrm{d}\omega_{k_1}\,
  \mathrm{d}\theta_{k_1}\; B_s(\kv_1,\kv_2)\, Y_L^M(\omega_{k_1},\theta_{k_1})\,.
\end{equation}
Restricting to the monopole $L=0$ and expanding the internal angle
$\mu_{12} \equiv \hat{\kv}_1\cdot\hat{\kv}_2$ in Legendre polynomials,
\begin{equation}
  B_{s,0}(k_1,k_2;\mu_{12})
  = \sum_{\ell=0}^{\ell_{\max}} B_{s,0,\ell}(k_1,k_2)\,\mathcal{L}_\ell(\mu_{12})\,,
\end{equation}
with
\begin{equation}
  B_{s,0,\ell}(k_1,k_2) \equiv \frac{2\ell+1}{2}
  \int_{-1}^{1} \mathrm{d}\mu_{12}\;
  B_{s,0}(k_1,k_2;\mu_{12})\,\mathcal{L}_\ell(\mu_{12})\,,
\end{equation}
and $k_3$ fixed by triangle closure; we use $\ell_{\max} = 10$.

\paragraph{3PCF.}
The analogous decomposition in configuration space proceeds identically.
The 3PCF is first expanded in spherical harmonics with respect to the
orientation of $\rv_{12}$ relative to $\nv$,
\begin{equation}
  \zeta_s(\rv_{12},\rv_{13}) = \sum_{L,M}
  \zeta_{s,L}^M(r_{12},r_{13},\mu_{12,13})\,
  Y_L^M(\omega_{r_{12}},\theta_{r_{12}})\,,
\end{equation}
where
\begin{equation}
  \zeta_{s,L}^M(r_{12},r_{13},\mu_{12,13}) \equiv
  \int \mathrm{d}\omega_{r_{12}}\,\mathrm{d}\theta_{r_{12}}\;
  \zeta_s(\rv_{12},\rv_{13})\,
  Y_L^{M*}(\omega_{r_{12}},\theta_{r_{12}})\,.
\end{equation}
Restricting to $L=0$ and expanding the internal angle
$\mu_{12,13} \equiv \hat{\rv}_{12}\cdot\hat{\rv}_{13}$ in Legendre
polynomials,
\begin{equation}
  \zeta_{s,0}(r_{12},r_{13};\mu_{12,13})
  = \sum_{\ell=0}^{\ell_{\max}}
  \zeta_{s,0,\ell}(r_{12},r_{13})\,\mathcal{L}_\ell(\mu_{12,13})\,,
\end{equation}
with
\begin{equation}
  \zeta_{s,0,\ell}(r_{12},r_{13}) \equiv \frac{2\ell+1}{2}
  \int_{-1}^{1} \mathrm{d}\mu_{12,13}\;
  \zeta_{s,0}(r_{12},r_{13};\mu_{12,13})\,
  \mathcal{L}_\ell(\mu_{12,13})\,.
\end{equation}
For simplicity, in the following we refer to the multipoles of the 
isotropic component in redshift space as, respectively, $B_{\ell}$ 
and $\zeta_{\ell}$.

\subsection{Redshift-space perturbation theory kernels}
\label{app:kernels}

We collect here the explicit expressions for the redshift-space kernels
$Z_n$ that appear in the perturbative expansion of
Eq.~\eqref{eq:delta_s_Zn}. These encode both the galaxy bias and the
velocity field contributions to the galaxy density contrast in redshift
space. The first-order kernel reads
\begin{equation}
  Z_1(\kv_1) = b_1 + f\mu_1^2\,,
  \label{eq:Z1}
\end{equation}
which reduces to the standard Kaiser factor \citep{Kaiser1987} when
evaluated on the total wavevector, $Z_1(\kv) = b_1 + f\mu^2$. The
second-order kernel is
\begin{align}
  Z_2(\kv_1,\kv_2) = &
  K_2(\kv_1,\kv_2)
  + f\mu^2\,G_2(\kv_1,\kv_2) \notag \\ &
  + \frac{f\mu k}{2}
  \left[
    \frac{\mu_1}{k_1}\bigl(b_1 + f\mu_2^2\bigr)
    +
    \frac{\mu_2}{k_2}\bigl(b_1 + f\mu_1^2\bigr)
  \right],
  \label{eq:Z2}
\end{align}
where $\mu \equiv \nv\cdot\hat{\kv}$ with $\kv = \kv_1+\kv_2$,
$\mu_i \equiv \nv\cdot\hat{\kv}_i$, and $k \equiv |\kv_1+\kv_2|$.
The bias kernel $K_2$ and the velocity kernel $G_2$ are defined as
\begin{align}
  K_2(\kv_1,\kv_2) &= \frac{b_2}{2}
  + b_1\,F_2(\kv_1,\kv_2)
  + b_{G_2}\,S(\kv_1,\kv_2)\,,
  \label{eq:K2}\\
  G_2(\kv_1,\kv_2) &= G_2^{\rm PT}(\kv_1,\kv_2)\,,
  \label{eq:G2}
\end{align}
where $F_2$ and $G_2$ are the standard SPT density and velocity kernels,
$S(\kv_1,\kv_2) \equiv (\hat{\kv}_1\cdot\hat{\kv}_2)^2 - 1/3$ is the
tidal shear invariant, and $b_{G_2}$ is the non-local bias parameter
associated with $\mathcal{G}_2$. The third-order kernel, which enters
the one-loop power spectrum and must be fully symmetrised over its
arguments, reads
\begin{align}
  Z_3(\kv_1,\kv_2,\kv_3) &=
  K_3(\kv_1,\kv_2,\kv_3)
  \notag \\ &\quad
  + f\mu^2\,G_3(\kv_1,\kv_2,\kv_3)
  + \frac{f^2\mu^2 k^2}{2}
  \bigl(b_1+f\mu_1^2\bigr)\frac{\mu_2\mu_3}{k_2 k_3}
  \notag\\
  &\quad
  + f\mu k\,\frac{\mu_3}{k_3}
  \Bigl[b_1\,F_2(\kv_1,\kv_2)
  + f\mu_{12}^2\,G_2(\kv_1,\kv_2)\Bigr]
  \notag\\
  &\quad
  + f\mu k\,\bigl(b_1+f\mu_1^2\bigr)
  \frac{\mu_{23}}{k_{23}}\,G_2(\kv_2,\kv_3)
  \notag\\
  &\quad
  + \frac{b_2}{2}\,f\mu k\,\frac{\mu_1}{k_1}
  + b_{G_2}\,f\mu k\,\frac{\mu_1}{k_1}\,S(\kv_2,\kv_3)\,,
  \label{eq:Z3}
\end{align}
where $\kv = \kv_1+\kv_2+\kv_3$, $\mu \equiv \nv\cdot\hat{\kv}$,
$k_{ij} \equiv |\kv_i+\kv_j|$ and
$\mu_{ij} \equiv \nv\cdot(\kv_i+\kv_j)/k_{ij}$.
The third-order bias kernel is
\begin{align}
  K_3(\kv_1,\kv_2,\kv_3) = &
  b_1\,F_3(\kv_1,\kv_2,\kv_3)
  + b_2\,F_2(\kv_1,\kv_2)
  \notag \\ &
  + b_{G_2}\Bigl[
    G_2(\kv_1,\kv_2)
    + F_2(\kv_1,\kv_2)\,S(\kv_1{+}\kv_2,\kv_3)
  \Bigr]
  \notag \\ &
  + b_{\Gamma_3}\,\gamma_3(\kv_1,\kv_2,\kv_3)\,,
  \label{eq:K3}
\end{align}
where $F_3$ is the third-order SPT density kernel, $G_3$ is the
corresponding velocity kernel, and $\gamma_3$ is the third-order tidal
operator associated with the bias parameter $b_{\Gamma_3}$. The full $Z_3$ kernel is obtained by summing over
all $3! = 6$ permutations of $(\kv_1,\kv_2,\kv_3)$ and dividing by $6$.

\section{Clustering estimators and measurements}
\label{app:clustering}

\subsection{Clustering estimators}
\label{app:estimators}

In what follows we describe the 2PCF and 3PCF estimators adopted, the harmonic decompositions employed, and the specific choices made for the data and mock analyses. Throughout this work we adopt the plane-parallel approximation. For the 2PCF, the line of sight (LOS) is defined for each pair of galaxies as the direction of its midpoint. For the 3PCF, we do not consider any explicit LOS dependence, since we consider only the LOS-averaged statistic. Both estimators belong to the class of Szapudi--Szalay estimators \citep{SzapudiSzalay1997}, which provide unbiased, minimum-variance counts for $N$-point statistics and naturally account for boundary and selection effects through the inclusion of random catalogues; in compact form, they can be written as $(D-R)^N/R^N$.

\subsubsection{Two-point correlation function}
\label{app:2pcf_estimator}

We estimate the anisotropic 2PCF using the Landy--Szalay estimator \citep{LandySzalay1993},
\begin{equation}
  \tilde{\xi}(s,\mu) =
  \frac{DD(s,\mu) - 2\,DR(s,\mu) + RR(s,\mu)}{RR(s,\mu)}\,,
  \label{eq:ls}
\end{equation}
where $DD(s,\mu)$, $DR(s,\mu)$ and $RR(s,\mu)$ are the normalised counts of data--data, data--random and random--random pairs in bins of separation $s$ and cosine angle $\mu = \hat{\mathbf{s}}\cdot\hat{\mathbf{n}}$ with respect to the LOS $\hat{\mathbf{n}}$. The estimator is applied independently to each of the four BOSS data chunks, and identically to each mock realisation. Each pair count is weighted by the product of the total weights $w_{\rm tot}$ of the two objects.

We compress the anisotropic 2PCF by projecting onto Legendre polynomials,
\begin{equation}
  \tilde{\xi}_\ell(s) = \frac{2\ell+1}{2}
  \int_{-1}^{1} d\mu\;\tilde{\xi}(s,\mu)\,\mathcal{L}_{\ell}(\mu)\,,
  \label{eq:xi_multipoles}
\end{equation}
and retain the monopole ($\ell=0$), quadrupole ($\ell=2$) and hexadecapole ($\ell=4$). The adopted binning is $\Delta s = 5\,h^{-1}\mathrm{Mpc}$ in separation and $\Delta\mu = 0.01$ for the cosine angle, with measurements reported at bin centres. The scales used in the cosmological analysis are $s \in [30,\,130]\,h^{-1}\mathrm{Mpc}$, discussed further in Sec.~\ref{sec:likelihood}.

\subsubsection{Three-point correlation function}
\label{app:3pcf_estimator}

The 3PCF can be estimated using the unbiased, minimum-variance
\citet{SzapudiSzalay1997} estimator,
\begin{equation}
  \tilde{\zeta}(\mathbf{x}_1,\mathbf{x}_2,\mathbf{x}_3)
  = \frac{DDD - 3\,DDR + 3\,DRR - RRR}{RRR}\,,
  \label{eq:ss3}
\end{equation}
where $DDD$, $DDR$, $DRR$, and $RRR$ denote the normalised triplet counts with
zero, one, two, and three random points, respectively. This estimator
automatically subtracts the disconnected part of the three-point statistics and
accounts for boundary and selection effects through the inclusion of the random
catalogue, making it well suited for a masked survey such as BOSS DR12. As for
the 2PCF, the estimator is applied identically to each mock catalogue, using the
associated random catalogue.

The brute-force evaluation of Eq.~\eqref{eq:ss3} scales as $\mathcal{O}(N^3)$
with the number of objects $N$, making it prohibitive for samples of the size of
BOSS DR12. To reduce the computational cost, we use the \textsc{MeasCorr} code
\citep{FarinaEtal2024a, EuclidGuidiEtAl2025, EuclidVeropalumbo2026}, which
implements the Spherical Harmonic Decomposition (SHD) algorithm presented in
\citet{SlepianEisenstein2015B, SlepianEisenstein2018}, achieving
$\mathcal{O}(N^2)$ scaling. The speed-up of the SHD algorithm is achieved by noting that, centred on each
galaxy $\mathbf{x}$, the radially binned density field can be expanded in
spherical harmonics,
\begin{equation}
  a_{\ell m}(r;\mathbf{x})
  = \int d\Omega\;\bar\delta(r,\hat{\mathbf{r}};\mathbf{x})\,Y^*_{\ell m}(\hat{\mathbf{r}})\,,
  \label{eq:alm}
\end{equation}
where $\bar\delta(r,\hat{\mathbf{r}};\mathbf{x})$ is the density field radially
binned in the shell of radius $r$ around the centre $\mathbf{x}$. Using the
spherical-harmonic addition theorem, the Legendre polynomial of the opening
angle between two side vectors $\hat{\mathbf{r}}_1$ and $\hat{\mathbf{r}}_2$
can be factored into a product of separately computed spherical-harmonic
coefficients,
\begin{equation}
  \mathcal{L}_{\ell}(\hat{\mathbf{r}}_1\cdot\hat{\mathbf{r}}_2)
  = \frac{4\pi}{2\ell+1}\sum_{m=-\ell}^{\ell}
    Y_{\ell m}(\hat{\mathbf{r}}_1)\,Y^*_{\ell m}(\hat{\mathbf{r}}_2)\,.
  \label{eq:addition_theorem}
\end{equation}
The local estimate of the $\ell$-th multipole of the 3PCF about the centre
$\mathbf{x}$ therefore becomes
\begin{equation}
  \hat{\bar\zeta}^{\rm loc}_\ell(r_1,r_2;\mathbf{x})
  = \frac{\delta(\mathbf{x})}{4\pi}
  \sum_{m=-\ell}^{\ell} a_{\ell m}(r_1;\mathbf{x})\,a^*_{\ell m}(r_2;\mathbf{x})\,,
  \label{eq:zeta_local}
\end{equation}
and the global multipole coefficient is obtained by averaging over all galaxies
acting as centres,
\begin{equation}
  \bar\zeta_\ell(r_1,r_2)
  = \frac{1}{N}\sum_{\mathbf{x}\in{\rm gal}}
  \hat{\bar\zeta}^{\rm loc}_\ell(r_1,r_2;\mathbf{x})\,.
  \label{eq:zeta_global}
\end{equation}
At any given centre, the $a_{\ell m}$ coefficients are precomputed once for each
radial bin, so no $\mathcal{O}(N^2)$ operation per centre is ever required; the
scaling reduces to $\mathcal{O}(N\times nV_{R_{\max}})$, i.e.\ $\mathcal{O}(N^2)$
overall, with the same favorable scaling as the two-point estimator.

In this work we focus on the isotropic component of the 3PCF with respect to
the line of sight, i.e.\ we consider only the monopole with resepect to the LOS,
which is equivalent to averaging the 3PCF over all orientations of the triangle
with respect to the LOS. The full LOS-anisotropic 3PCF is left for future work.
We retain multipoles up to $\ell_{\max} = 4$, which capture the bulk of the
cosmological constraining power of the 3PCF while keeping the data-vector
dimension manageable. Including higher multipoles would in principle add
information, but at the cost of inflating the data vector, degrading the
conditioning of the joint covariance matrix and thereby reducing the stability
of its inversion and the reliability of the resulting parameter inference. The
choice $\ell_{\max} = 4$ therefore represents a principled optimum between
information content and numerical stability of the likelihood analysis.

\subsection{Clustering measurements}
\label{app:measurements}

\begin{figure*}[t!]
  \centering
  \includegraphics[width=\textwidth]{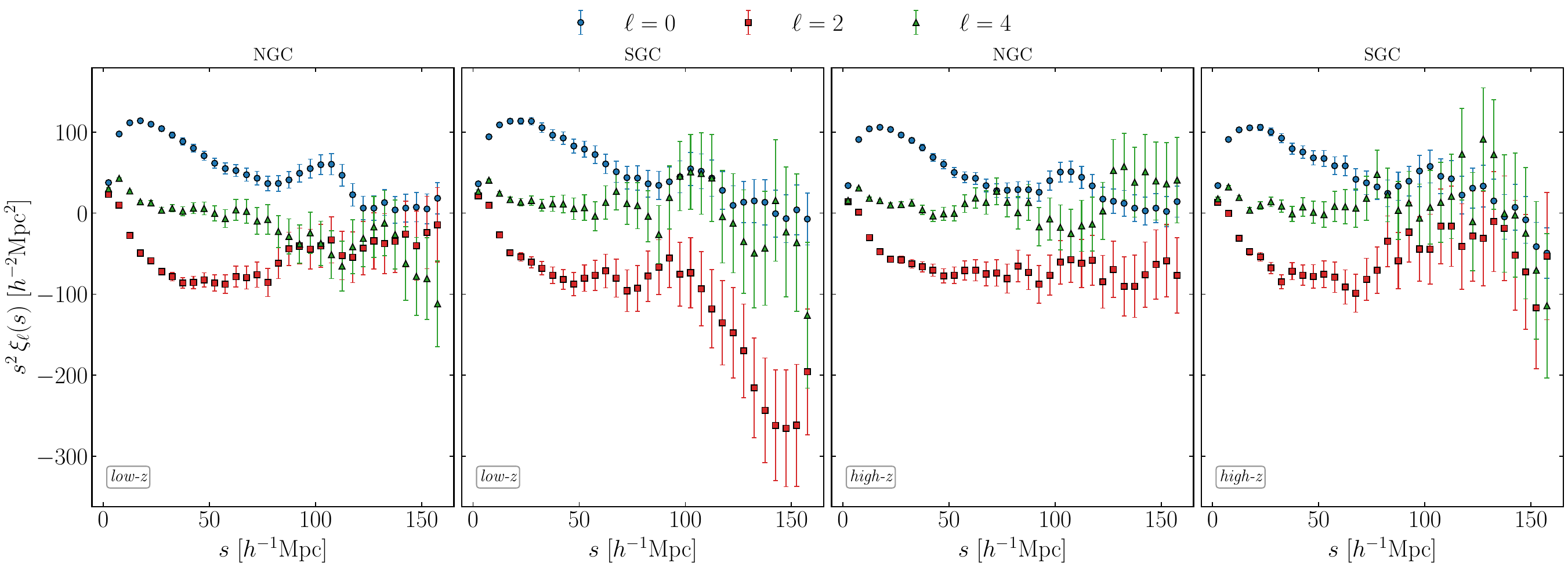}
  \caption{Multipoles of the two-point correlation function measured from
  the four BOSS DR12 data chunks. Each panel shows $s^2\,\xi_\ell(s)$ as
  a function of separation $s$, for the monopole ($\ell=0$, blue),
  quadrupole ($\ell=2$, orange) and hexadecapole ($\ell=4$, green).
  Columns correspond to the two sky regions (NGC, left; SGC, right),
  while rows correspond to the two redshift bins (\textit{low-z} with
  $z_{\rm eff}\simeq0.32$, top; \textit{high-z} with
  $z_{\rm eff}\simeq0.57$, bottom). Measurements are performed in the
  range $s \in [0, 150]\,h^{-1}{\rm Mpc}$ with bin width
  $\Delta s = 5\,h^{-1}{\rm Mpc}$; bin centres are located at
  $s = 2.5, 7.5, \ldots, 147.5\,h^{-1}{\rm Mpc}$. Error bars are the
  square root of the diagonal of the 2PCF covariance matrix
  $\hat{C}_{\xi\xi}$, estimated from the 2048 MultiDark-Patchy mock
  catalogues.}
  \label{fig:2pcf_meas}
\end{figure*}

\begin{figure*}[t!]
    \centering
    \includegraphics[width=1.0\textwidth]{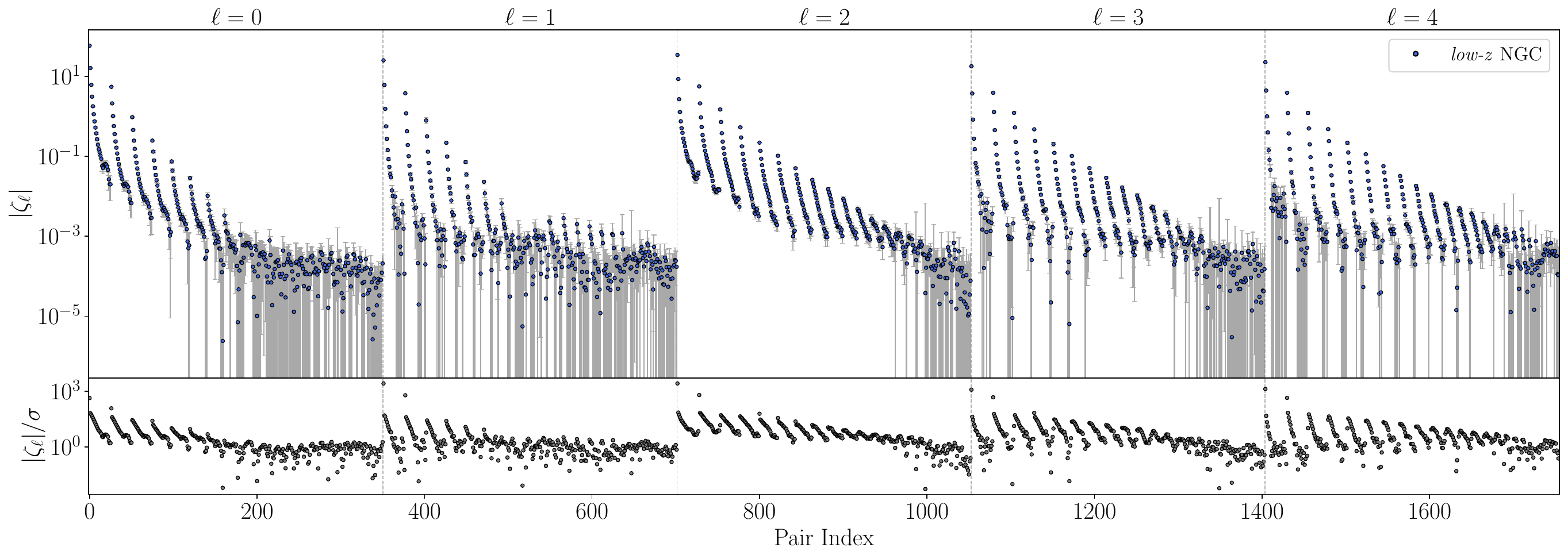}
    \caption{Multipoles of the three-point correlation function measured from
    the NGC-\textit{low-z} chunk of BOSS DR12 ($z_{\rm eff}\simeq0.32$),
    shown as a function of the ordered pair index $(r_1, r_2)$ with
    $r_1 \leq r_2$. Triangles are measured over the range
    $r_{1,2}\in[0,150]\,h^{-1}{\rm Mpc}$ with bin width
    $\Delta r = 5\,h^{-1}{\rm Mpc}$; pairs are sorted in ascending order of
    $r_1$ and then $r_2$. \textit{Top panel:} measured multipoles
    $\zeta_\ell(r_1,r_2)$ for $\ell = 0, 1, 2, 3, 4$; grey error bars are
    the square root of the diagonal of $\hat{C}_{\zeta\zeta}$, estimated
    from the 2048 Patchy mocks. \textit{Bottom panel:} absolute
    signal-to-noise ratio $|\zeta_\ell(r_1,r_2)|/\sigma_\zeta(r_1,r_2)$
    for each multipole and configuration.}
    \label{fig:3pcf_meas}
\end{figure*}
We present here the measurements of the 2PCF and 3PCF extracted from the four BOSS DR12 data chunks (NGC/SGC $\times$ \textit{low-z}/\textit{high-z}) using the estimators described in Sec.~\ref{app:estimators}. In both cases, error bars on individual data points represent the square root of the diagonal of the respective covariance matrix block, estimated from the full suite of 2048 MultiDark-Patchy mock catalogues as described in Sec.~\ref{sec:covariance}. The fiducial cosmological parameters of the mock catalogues, used to convert angles and redshifts into comoving coordinates, are listed in Tab.~\ref{tab:patchy_cosmo}.

\subsubsection{2PCF measurements}
\label{app:2pcf_measurements}

The anisotropic 2PCF $\tilde{\xi}(s,\mu)$ is measured for each of the four data chunks using the Landy--Szalay estimator of Eq.~\eqref{eq:ls}. We bin pair separations over the range $s \in [0, 150]\,h^{-1}{\rm Mpc}$ with a fixed bin width of $\Delta s = 5\,h^{-1}{\rm Mpc}$, yielding 30 bins with centres at $s = 2.5, 7.5, \ldots, 147.5\,h^{-1}{\rm Mpc}$. The anisotropic signal is then compressed into the monopole ($\ell = 0$), quadrupole ($\ell = 2$) and hexadecapole ($\ell = 4$) multipole moments via Eq.~\eqref{eq:xi_multipoles}.

The resulting measurements, multiplied by $s^2$ to enhance the visibility of the clustering signal, are shown in Fig.~\ref{fig:2pcf_meas} for all four chunks. The monopole is detected with high signal-to-noise across the full separation range, and shows the characteristic bump at $s \sim 100\,h^{-1}{\rm Mpc}$ associated with the baryon acoustic oscillation (BAO) feature. The measurements from the NGC and SGC regions are mutually consistent within the errors, as expected for a survey with homogeneous selection. The \textit{high-z} sample shows a systematically larger clustering amplitude compared to \textit{low-z}, reflecting the higher effective linear bias of the more massive galaxy population targeted by the CMASS selection at $z_{\rm eff} \simeq 0.57$.

\subsubsection{3PCF measurements}
\label{app:3pcf_measurements}

The 3PCF multipoles $\zeta_\ell(r_1, r_2)$ for $\ell = 0, 1, 2, 3, 4$ are measured for each data chunk using the \textsc{MeasCorr} package \citep{FarinaEtal2024a}, which implements the SHD estimator described in App. ~\ref{app:3pcf_estimator}. The two side lengths are binned over the range $r_{1,2} \in [0, 150]\,h^{-1}{\rm Mpc}$ with $\Delta r = 5\,h^{-1}{\rm Mpc}$, yielding bin centres at $2.5, 7.5, \ldots, 147.5\,h^{-1}{\rm Mpc}$. Triangle configurations are parametrised by ordered pairs $(r_1, r_2)$ with $r_1 \leq r_2$, avoiding double counting; the pairs are sorted first in ascending order of $r_1$ and then of $r_2$, defining a unique Pair Index running over all $N_{\rm pairs} = N_{\rm bins}(N_{\rm bins}+1)/2$ configurations.

Figure~\ref{fig:3pcf_meas} shows the NGC measurements for the \textit{low-z} bin; the SGC results are fully consistent and enter the analysis on equal footing. The monopole ($\ell = 0$) dominates in amplitude across the full range of configurations, while the higher multipoles ($\ell = 1, 2, 3, 4$) are detected with $|\zeta_\ell|/\sigma > 1$ over a significant fraction of configurations at small scales. At large separations all multipoles fall below the noise level, confirming that the effective information content of the 3PCF is dominated by the small-scale configurations retained in the data vector.

\begin{table}
  \centering
  \caption[Cosmological parameters of MultiDark-Patchy mocks]
  {Fiducial cosmological parameters of the MultiDark-Patchy mock catalogues
  \citep{KitauraEtal2016, Klypin2016}.}
  \label{tab:patchy_cosmo}
  \begin{tabular}{lc}
    \hline\hline
    Parameter & Value \\
    \hline
    $\Omega_m$         & 0.307115              \\
    $\Omega_\Lambda$   & 0.692885              \\
    $\Omega_b$         & 0.048                 \\
    $\sigma_8$         & 0.8288                \\
    $h$                & 0.6777                \\
    $n_s$              & 0.9611                \\
    $A_s$              & $2.106\times10^{-9}$  \\
    $M_\nu$            & 0                     \\
    $\Omega_k$         & 0                     \\
    \hline\hline
  \end{tabular}
\end{table}

\section{Dependence of the 2PCF inference on the minimum scale cut}
\label{app:scale_cut_2pcf}

This appendix explores the dependence of the cosmological 
inference on the minimum scale cut $r_{\rm min}$, varying between
$20$ and $80\,h^{-1}{\rm Mpc}$. The goal is to assess the 
stability of the posterior distributions and the goodness of fit 
as progressively smaller scales are included in the analysis.

Table~\ref{tab:2pcf_corrections} reports the data-vector size 
$N_d$, the number of degrees of freedom $N_\mathrm{dof} = N_d 
- N_p$, and the Hartlap and Percival correction factors as a 
function of $r_\mathrm{min}$. Since $N_m = 2048 \gg N_d$ 
across all scale cuts, the Hartlap correction remains within 
$\lesssim 3\%$ of unity, while the Percival correction amounts 
to at most $\sim 12\%$ at $r_\mathrm{min} = 20\,h^{-1}$Mpc, 
decreasing to $\sim 4\%$ at $r_\mathrm{min} = 80\,h^{-1}$Mpc. 
Both corrections are properly accounted for in the likelihood 
evaluation.

\begin{table}
\centering
\caption{Data-vector size and covariance correction factors 
for the 2PCF-only analysis as a function of $r_\mathrm{min}$, 
with $r_\mathrm{max} = 130\,h^{-1}$Mpc fixed. $N_d$ is the 
number of data points, $N_\mathrm{dof} = N_d - N_p$ the 
degrees of freedom, and $\alpha_\mathrm{H}$, $\alpha_\mathrm{P}$ 
the Hartlap and Percival correction factors 
(Eqs.~\ref{eq:hartlap}--\ref{eq:percival}).}
\label{tab:2pcf_corrections}
\begin{tabular}{ccccc}
\hline\hline
$r_\mathrm{min}\ [h^{-1}\mathrm{Mpc}]$ & $N_d$ & 
$\alpha_\mathrm{H}$ & $\alpha_\mathrm{P}$ & $N_\mathrm{dof}$ \\
\hline
20 & 264 & 0.9711 & 1.1249 & 245 \\
30 & 240 & 0.9734 & 1.1101 & 221 \\
40 & 216 & 0.9761 & 1.0957 & 197 \\
50 & 192 & 0.9784 & 1.0817 & 173 \\
60 & 168 & 0.9812 & 1.0680 & 149 \\
70 & 144 & 0.9839 & 1.0547 & 125 \\
80 & 120 & 0.9863 & 1.0416 & 101 \\
\hline
\end{tabular}
\end{table}

\begin{figure}
  \includegraphics[width=1\columnwidth]{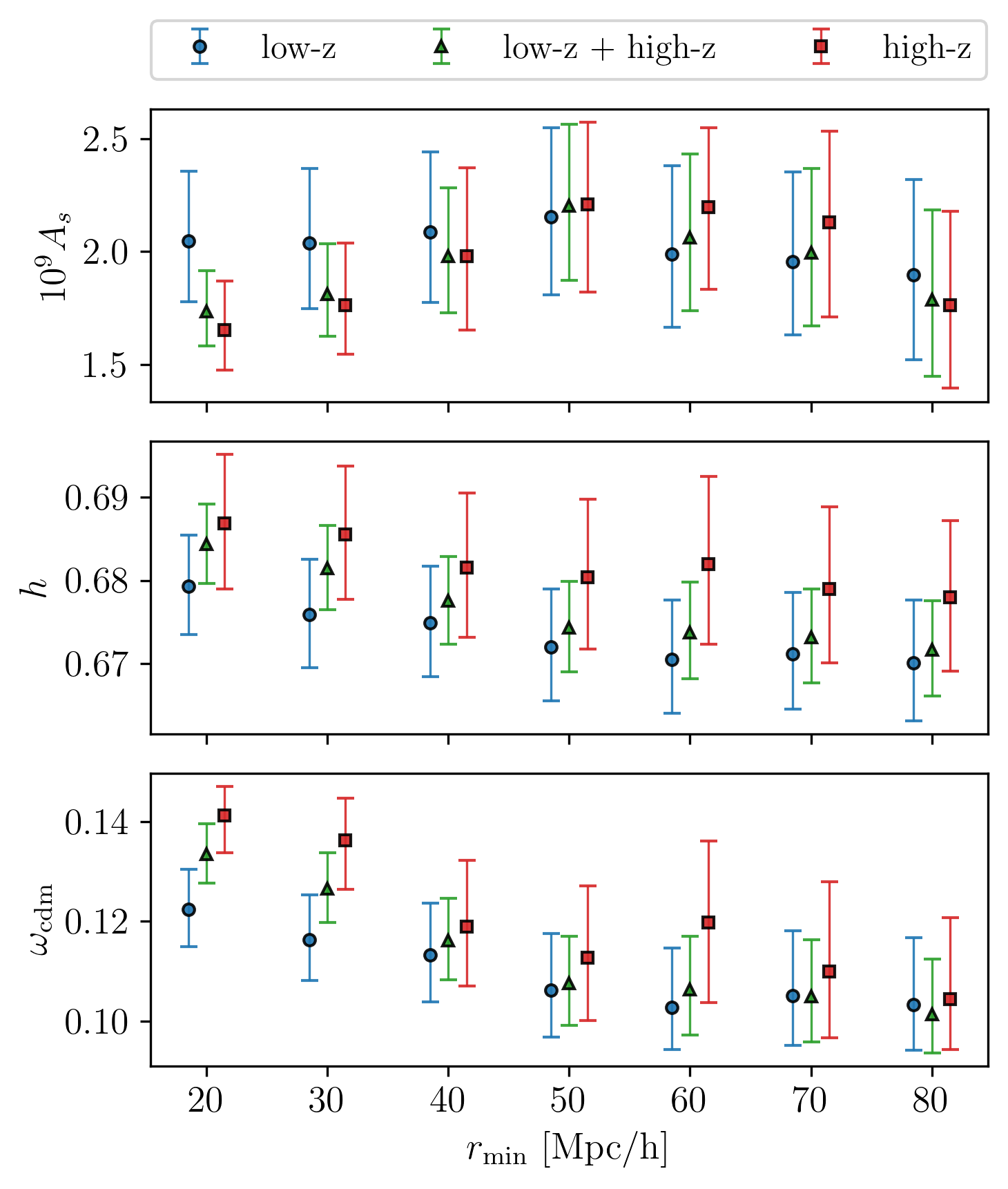}
  \caption{Marginalised posterior means and $68\%$ credible intervals for
  $10^9 A_s$ (top), $h$ (middle), and $\omega_{\rm cdm}$ (bottom) as a
  function of the minimum fitting scale $r_{\rm min}$ for the
  \textit{low-z} (blue circles), \textit{high-z} (red squares), and
  \textit{combined} (green triangles) 2PCF analyses. The grey dashed
  horizontal line marks the fiducial value of each parameter.}
  \label{fig:cosmo_vs_rmin}
\end{figure}

\paragraph{Cosmological parameters.}
Figure~\ref{fig:cosmo_vs_rmin} shows the dependence of the marginalised
posterior means and $68\%$ credible intervals of $10^9 A_s$, $h$, and
$\omega_{\rm cdm}$ as a function of $r_{\rm min}$ for the
\textit{low-z}, \textit{high-z}, and \textit{combined} analyses.

The Hubble parameter $h$ is remarkably stable across the full range of
scale cuts, with no evidence of a significant trend and excellent agreement
between the two redshift bins at all values of $r_{\rm min}$. The scalar amplitude $10^9 A_s$ shows a mild dependence on $r_{\rm min}$,
with a gradual increase in its posterior mean and a corresponding broadening
of its uncertainty as smaller scales are progressively removed from the
analysis. This behaviour reflects the loss of small-scale information that
contributes to constraining the overall amplitude of matter fluctuations. The cold dark matter density $\omega_{\rm cdm}$ exhibits a more pronounced
dependence on $r_{\rm min}$, with systematically lower values favoured when
smaller scales are included. This trend is consistent with an evolving
degeneracy between $A_s$ and $\omega_{\rm cdm}$ as broad-band shape
information is progressively added or removed from the data vector. The
\textit{low-z} and \textit{high-z} samples remain broadly consistent across
the full range of scale cuts, with differences typically at the
$\sim 1$--$2\sigma$ level at most.

\begin{figure}
  \includegraphics[width=.49\textwidth]{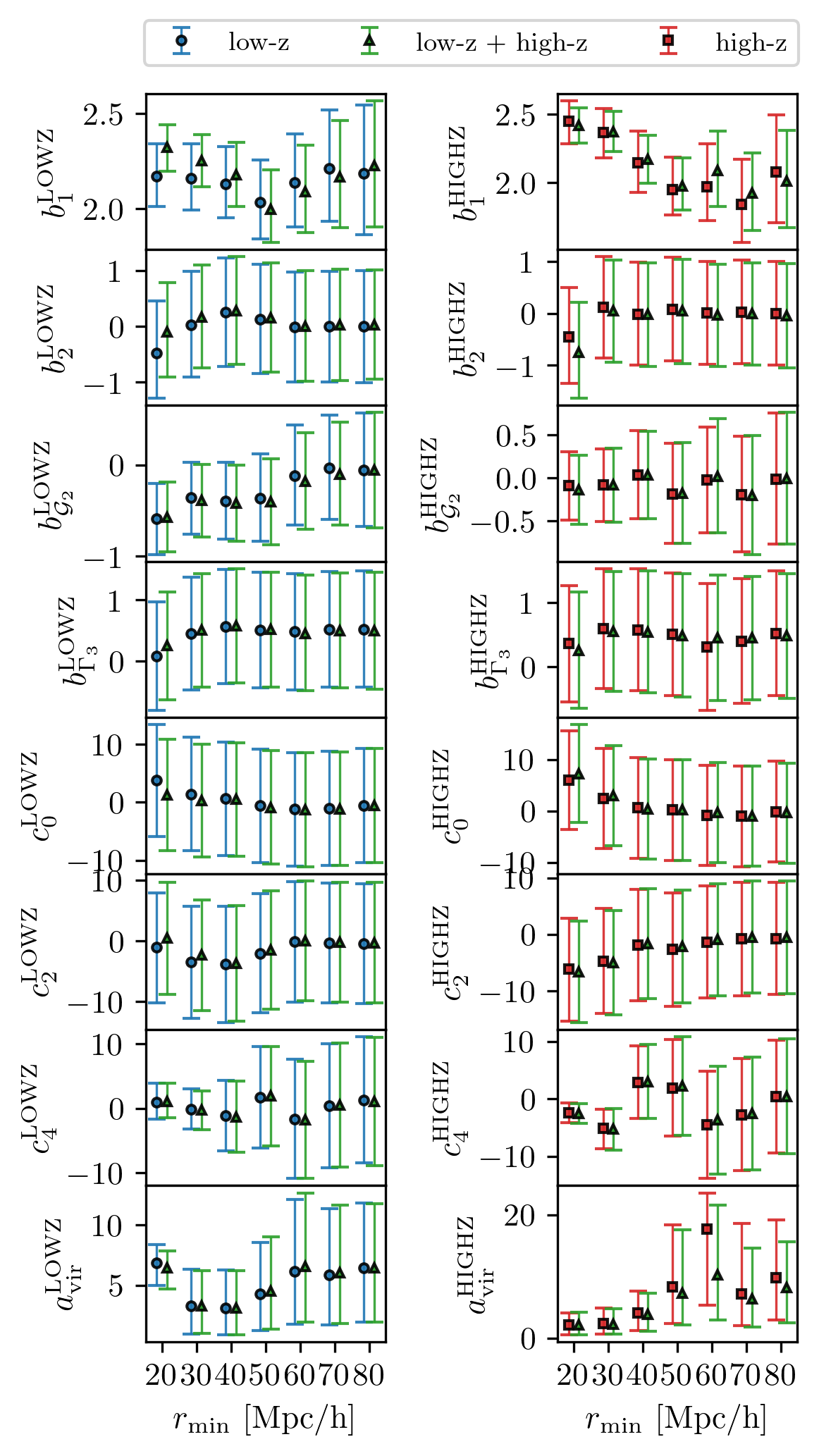}
  \caption{Marginalised posterior means and $68\%$ credible intervals for
  the nuisance parameters as a function of $r_{\rm min}$, for the
  \textit{low-z} bin (left column, blue circles) and the \textit{high-z}
  bin (right column, red squares); green triangles show the corresponding
  results from the \textit{combined} analysis.}
  \label{fig:bias_vs_rmin}
\end{figure}

\paragraph{Nuisance parameters.}
Figure~\ref{fig:bias_vs_rmin} shows the corresponding behaviour of the
nuisance parameters as a function of $r_{\rm min}$. The linear bias
$b_1$ and velocity dispersion parameter $a_{\rm vir}$ are stable across
all scale cuts and for both redshift bins, indicating robustness with
respect to the choice of minimum scale. In contrast, the EFT counterterms $c_0$ and $c_2$ exhibit a smooth
dependence on $r_{\rm min}$, with larger absolute values when smaller
scales are included in the analysis. This reflects their role in
absorbing residual non-linear contributions rather than indicating any
breakdown of the perturbative model. The higher-order bias parameters
remain consistent with zero within uncertainties for all scale cuts
considered.

However, the goodness of fit, quantified via the $\chi^2$ statistic, as shown in Fig.~\ref{fig:chi2_vs_rmin_2PCF}), remains statistically acceptable across
the entire range of scale cuts considered. No significant degradation of
the fit is observed when including smaller scales down to
$r_{\rm min} = 20\,h^{-1}{\rm Mpc}$, indicating that the model provides a
consistent description of the data within the explored range. Overall, the dependence of both cosmological and nuisance parameters on
$r_{\rm min}$ is smooth and well-behaved, with no indication of a sharp
transition in the validity of the model. In the absence of statistically
significant degradation in the $\chi^2$ statistic, we adopt
$r_{\rm min} = 20\,h^{-1}{\rm Mpc}$ as the fiducial choice for the baseline
analysis, as it maximises the available information content while remaining
fully consistent with the goodness-of-fit criterion.

\section{Scale cut and posterior validation for the joint analysis}
\label{app:scale_cut_joint}

We present here the supporting diagnostics for the joint 2PCF+3PCF analysis
of Sect.~\ref{sec:joint_analysis}, complementing the cosmological parameter
running and goodness-of-fit discussed in the main text with the behaviour
of the nuisance parameters and the full posterior distributions.Also table~\ref{tab:joint_corrections} reports, for each combination 
of $r^\mathrm{3PCF}_\mathrm{min}$ and $\eta_\mathrm{min}$, the 
total size of the joint data vector $N_d$, the number of degrees 
of freedom $N_\mathrm{dof} = N_d - N_p$ with $N_p = 19$, and 
the Hartlap and Percival correction factors 
(Eqs.~\ref{eq:hartlap}--\ref{eq:percival}), with 
$r^\mathrm{2PCF}_\mathrm{min} = 20\,h^{-1}$Mpc fixed throughout. 
Unlike the 2PCF-only case, where both corrections are negligible 
($\lesssim 0.6\%$), the ratio $N_d/N_m$ is non-negligible across 
all joint configurations: the Hartlap correction reduces the 
effective precision matrix by up to $\sim 22\%$ at the smallest 
scale cut, while the Percival correction inflates the marginalised 
parameter variances by up to $\sim 25\%$. Both corrections 
decrease monotonically as $r^\mathrm{3PCF}_\mathrm{min}$ 
increases and as $\eta_\mathrm{min}$ increases, reflecting the 
progressive reduction of the data-vector size.

\begin{figure}
  \includegraphics[width=0.5\textwidth]{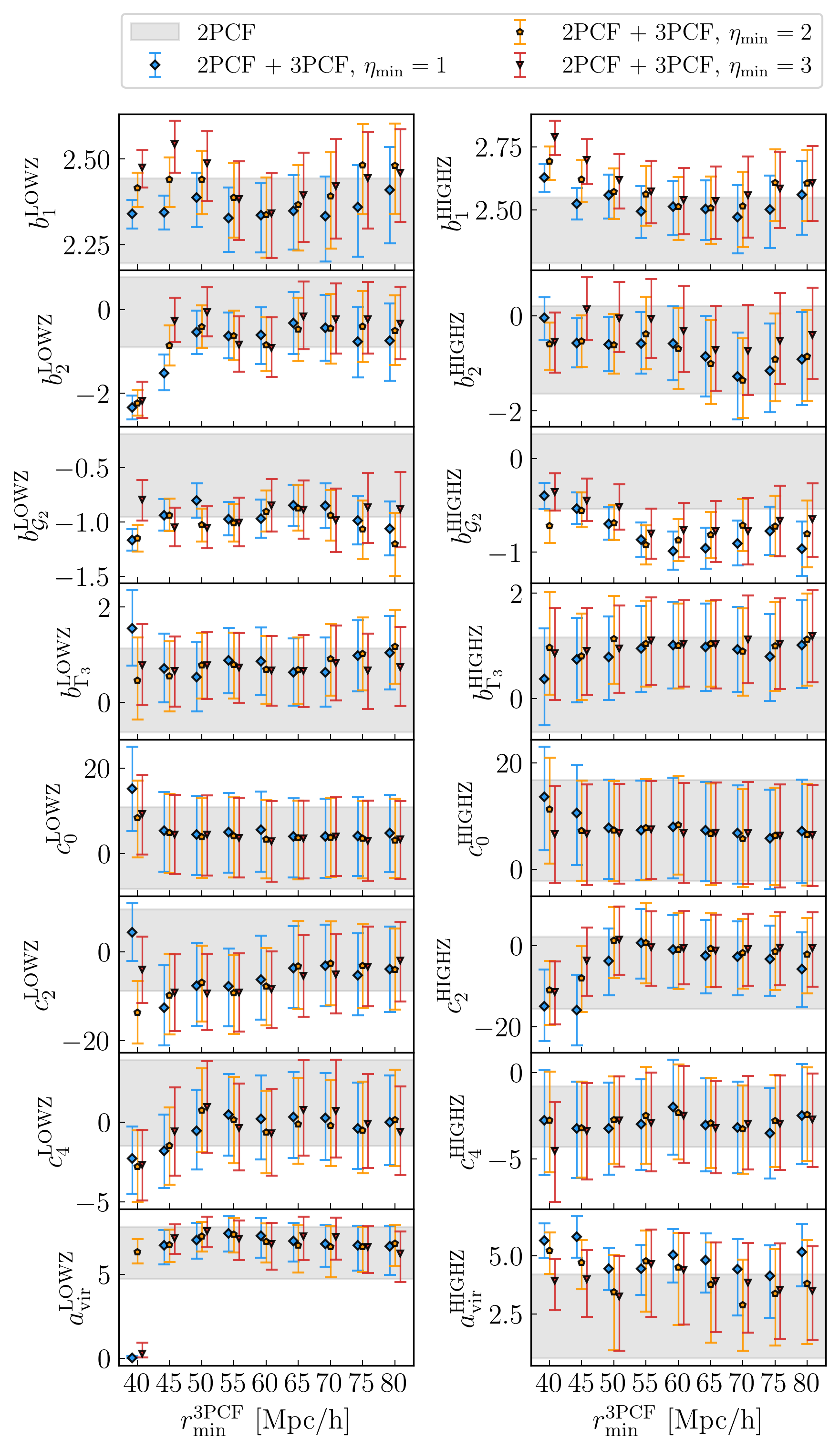}
  \caption{Marginalised posterior means and $68\%$ credible intervals for
  the nuisance parameters of the joint 2PCF+3PCF \textit{combined}
  analysis as a function of $r_{\rm min}^{\rm 3PCF}$, for the
  \textit{low-z} (left column) and \textit{high-z} (right column) bins.
  Symbols as in Fig.~\ref{fig:cosmo_vs_rmin3pcf_joint}: blue diamonds
  ($\eta_{\rm min} = 1$), green pentagons ($\eta_{\rm min} = 2$), orange
  triangles ($\eta_{\rm min} = 3$).}
  \label{fig:bias_vs_rmin3pcf_joint}
\end{figure}

\begin{table}
\centering
\caption{Joint data-vector size and covariance correction factors 
as a function of $r^\mathrm{3PCF}_\mathrm{min}$ and 
$\eta_\mathrm{min}$, with $r^\mathrm{2PCF}_\mathrm{min} = 
20\,h^{-1}$Mpc fixed. $N_d$ is the total number of data points 
in the original joint 2PCF$+$3PCF vector, $N_\mathrm{SV}^*$ the number of 
modes retained after covariance compression, $N_\mathrm{dof} = N_\mathrm{SV}^* 
- N_p$ the degrees of freedom with $N_p = 19$, and 
$\alpha_\mathrm{H}$, $\alpha_\mathrm{P}$ the Hartlap and Percival 
correction factors (Eqs.~\ref{eq:hartlap}--\ref{eq:percival}).}
\label{tab:joint_corrections}
\footnotesize
\begin{tabular}{ccccccc}
\hline\hline
$r^\mathrm{3PCF}_\mathrm{min}\ [h^{-1}\mathrm{Mpc}]$ & 
$\eta_\mathrm{min}$ & $N_d$ & $N_\mathrm{SV}^*$ &
$\alpha_\mathrm{H}$ & $\alpha_\mathrm{P}$ & $N_\mathrm{dof}$ \\
\hline
30 & 1 & 3304 & 433 & 0.7829 & 1.2487 & 414 \\
30 & 2 & 3000 & 393 & 0.8029 & 1.2180 & 374 \\
30 & 3 & 2712 & 366 & 0.8164 & 1.1981 & 347 \\
\hline
35 & 1 & 3000 & 396 & 0.8014 & 1.2202 & 377 \\
35 & 2 & 2712 & 358 & 0.8204 & 1.1923 & 339 \\
35 & 3 & 2440 & 333 & 0.8329 & 1.1746 & 314 \\
\hline
40 & 1 & 2712 & 360 & 0.8194 & 1.1937 & 341 \\
40 & 2 & 2440 & 326 & 0.8364 & 1.1697 & 307 \\
40 & 3 & 2184 & 304 & 0.8474 & 1.1547 & 285 \\
\hline
45 & 1 & 2440 & 331 & 0.8339 & 1.1732 & 312 \\
45 & 2 & 2184 & 301 & 0.8489 & 1.1527 & 282 \\
45 & 3 & 1944 & 280 & 0.8594 & 1.1387 & 261 \\
\hline
50 & 1 & 2184 & 304 & 0.8474 & 1.1547 & 285 \\
50 & 2 & 1944 & 279 & 0.8599 & 1.1381 & 260 \\
50 & 3 & 1720 & 260 & 0.8694 & 1.1258 & 241 \\
\hline
55 & 1 & 1944 & 286 & 0.8564 & 1.1427 & 267 \\
55 & 2 & 1720 & 261 & 0.8689 & 1.1264 & 242 \\
55 & 3 & 1512 & 244 & 0.8774 & 1.1156 & 225 \\
\hline
60 & 1 & 1720 & 270 & 0.8644 & 1.1322 & 251 \\
60 & 2 & 1512 & 244 & 0.8774 & 1.1156 & 225 \\
60 & 3 & 1320 & 227 & 0.8859 & 1.1050 & 208 \\
\hline
65 & 1 & 1512 & 256 & 0.8714 & 1.1232 & 237 \\
65 & 2 & 1320 & 230 & 0.8844 & 1.1069 & 211 \\
65 & 3 & 1144 & 216 & 0.8914 & 1.0983 & 197 \\
\hline
70 & 1 & 1320 & 244 & 0.8774 & 1.1156 & 225 \\
70 & 2 & 1144 & 219 & 0.8899 & 1.1001 & 200 \\
70 & 3 &  984 & 204 & 0.8974 & 1.0910 & 185 \\
\hline
75 & 1 & 1144 & 232 & 0.8834 & 1.1081 & 213 \\
75 & 2 &  984 & 208 & 0.8954 & 1.0934 & 189 \\
75 & 3 &  840 & 193 & 0.9030 & 1.0844 & 174 \\
\hline
80 & 1 &  984 & 222 & 0.8884 & 1.1019 & 203 \\
80 & 2 &  840 & 196 & 0.9015 & 1.0862 & 177 \\
80 & 3 &  712 & 182 & 0.9085 & 1.0779 & 163 \\
\hline\hline
\end{tabular}
\end{table}

\subsection*{Nuisance parameter stability}

Figure~\ref{fig:bias_vs_rmin3pcf_joint} shows the marginalised posterior
means and $68\%$ credible intervals for the nuisance parameters of both
redshift bins as a function of $r_{\rm min}^{\rm 3PCF}$, for the three
elongation cuts $\eta_{\rm min} \in \{1, 2, 3\}$.

The linear bias $b_1$ is stable across the entire range of $r_{\rm
min}^{\rm 3PCF}$ for both redshift bins and all elongation cuts, with no
appreciable systematic trend, confirming that this parameter is robustly
constrained independently of the 3PCF scale cut. The velocity-dispersion
parameter $a_{\rm vir}$ shows a notable anomaly at $r_{\rm min}^{\rm
3PCF} = 40\,h^{-1}{\rm Mpc}$ for $\eta_{\rm min} = 1$, where it drops
to near-zero values: this signals that nearly-degenerate triangle
configurations, admitted by the loose elongation cut, introduce squeezed
configurations in which the FoG damping is poorly constrained by the data.
This pathological behaviour disappears for $\eta_{\rm min} \geq 2$ and
vanishes entirely for all $\eta_{\rm min}$ from $r_{\rm min}^{\rm 3PCF}
\simeq 50\,h^{-1}{\rm Mpc}$ onward, providing an independent motivation
for the choice $\eta_{\rm min} = 3$ as the reference elongation cut.

The EFT counterterms $c_0$ and $c_2$ exhibit enhanced scatter and
systematically large absolute values at $r_{\rm min}^{\rm 3PCF} \leq
45\,h^{-1}{\rm Mpc}$, most pronounced for $\eta_{\rm min} = 1$. This
behaviour mirrors the instability seen in the cosmological parameters at
the same scale cuts (Fig.~\ref{fig:cosmo_vs_rmin3pcf_joint}) and is
consistent with the EFT counterterms absorbing non-perturbative
small-scale contributions beyond the range of validity of the model. From
$r_{\rm min}^{\rm 3PCF} \simeq 55\,h^{-1}{\rm Mpc}$ all three elongation
cuts yield counterterm values consistent with perturbative expectations,
with no residual trend. The higher-order bias parameters $b_2$,
$b_{G_2}$, $b_{\Gamma_3}$ and $c_4$ are consistent with zero within their
uncertainties throughout, with no significant dependence on $r_{\rm
min}^{\rm 3PCF}$ or $\eta_{\rm min}$.

Taken together with the $\chi^2_{\rm red}$ criterion
(Fig.~\ref{fig:chi2_fomgain}) and the cosmological parameter running
(Fig.~\ref{fig:cosmo_vs_rmin3pcf_joint}), the nuisance parameter
behaviour consistently identifies $r_{\rm min}^{\rm 3PCF} =
60\,h^{-1}{\rm Mpc}$ and $\eta_{\rm min} = 3$ as the conservative
threshold below which the perturbative model cannot be considered reliable.

\subsection*{Full posterior distributions}

\begin{figure*}
  \centering
  \includegraphics[width=\textwidth]{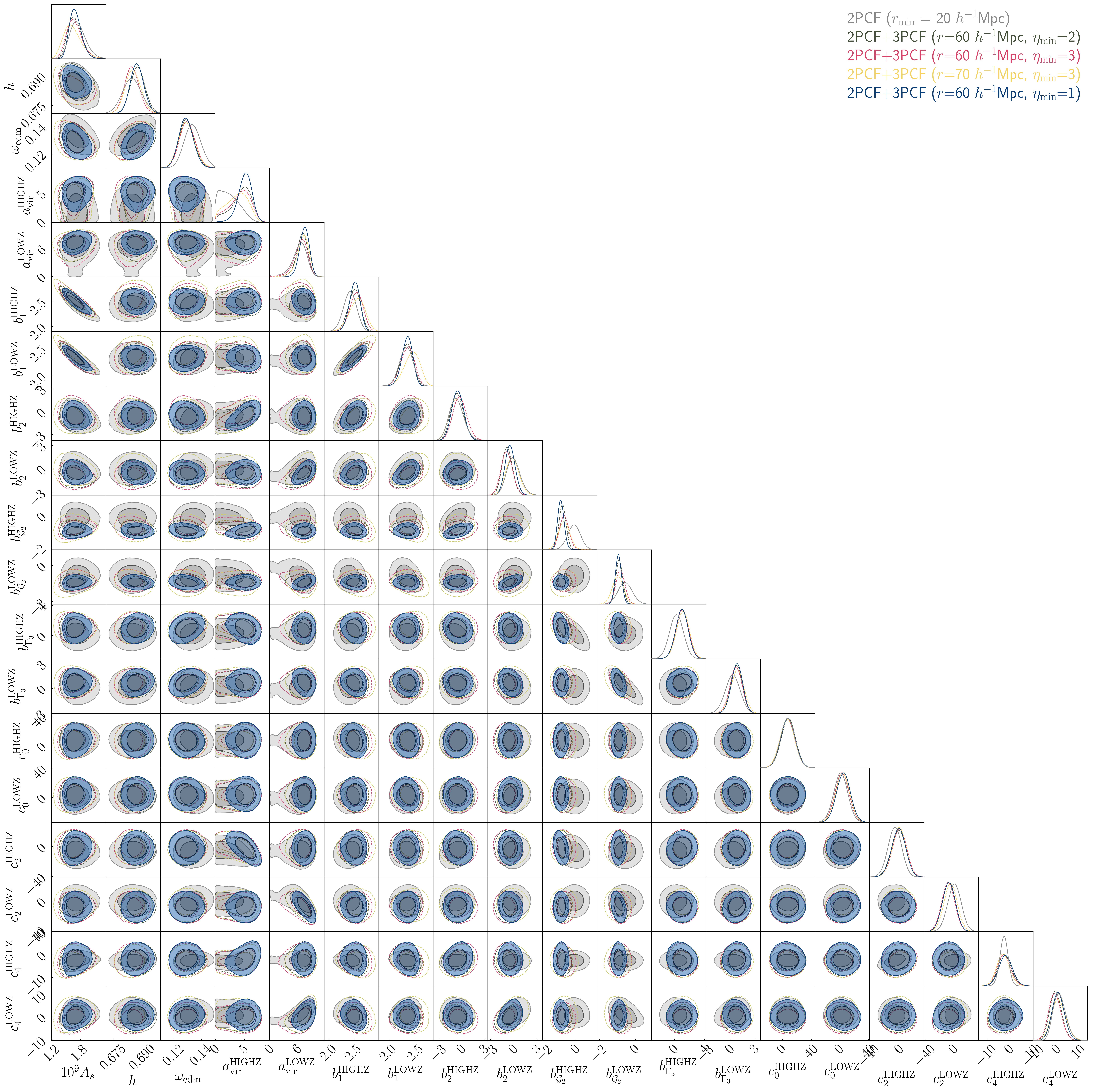}
  \caption{As in Fig. \ref{fig:chains_joint_cosmo}, but showing all the parameter set.}
  \label{fig:chain_joint_full}
\end{figure*}

Figure~\ref{fig:chain_joint_full} shows the complete marginalised
posterior distributions for the reference joint configuration ($r_{\rm
min}^{\rm 3PCF} = 60\,h^{-1}{\rm Mpc}$, $\eta_{\rm min} = 1$), extending
the three-parameter cosmological corner plot of
Fig.~\ref{fig:chains_joint_cosmo} to include all nuisance parameters for
both redshift bins.

Several inter-parameter correlations are visible in the full posterior.
The well-known degeneracy between $b_1$ and $10^9 A_s$ is clearly
present, as the two parameters jointly control the overall clustering
amplitude; this degeneracy is partially broken by the broad-band shape
information encoded in $\omega_{\rm cdm}$ and $h$, with a residual
negative correlation between $10^9 A_s$ and $\omega_{\rm cdm}$
consistent with their joint sensitivity to the power spectrum shape.
The nuisance parameters of the two redshift bins show no significant
cross-bin correlations, validating the assumption of independent bias
parameters in the combined analysis.

A comparison between the 2PCF-only and the joint 2PCF+3PCF
posteriors reveals a systematic tightening of the marginalised
constraints across the full parameter space. Beyond the cosmological
parameters discussed in Sect.~\ref{sec:joint_analysis}, the improvement
is most apparent in the linear bias parameters $b_1^{\rm LOWZ}$ and
$b_1^{\rm HIGHZ}$, reflecting the direct sensitivity of the 3PCF to
the large-scale bias through its leading-order dependence on the
squared density field. A tightening is also visible for $b_2$
and $b_{\mathcal{G}_2}$, consistent with the 3PCF's constraining power on higher-order bias parameters. The constraints on $a_{\rm vir}$ shift slightly but remain fully
consistent between the two analyses, suggesting that the 3PCF mildly
breaks the degeneracy between the finger-of-God damping scale and the
linear bias parameters through its sensitivity to the angular
structure of triangle configurations along the line of sight. We note 
that $a_{\rm vir}$ is additionally susceptible to prior volume effects 
arising from its degeneracies with $b_1$ and $10^9A_s$; however, since 
this parameter does not enter the cosmological inference directly, 
its marginalised posterior is not a primary diagnostic of this analysis.

\end{appendix}

\end{document}